\documentclass[aps,prd,twocolumn,amsmath,amssymb,showpacs,preprintnumbers,superscriptaddress,nofootinbib]{revtex4-1}

\usepackage{slashed}
\usepackage{bm} 
\usepackage{epsf}
\usepackage{graphicx,epsfig}

\usepackage{latexsym,amssymb,amsmath,float,url,mathrsfs}
\usepackage{bbold}
\usepackage{enumerate}
\usepackage{multirow,array,booktabs}
\usepackage{epstopdf}

\newcommand{\nc}{\newcommand}
\def\cal{\mathcal}
\def\rm{\mathrm}

\nc{\eq}[1]{Eq.~(\ref{#1})}
\nc{\eqs}[1]{Eqs.~(\ref{#1})}

\nc{\pd}{\partial}

\nc{\bea}{\begin{eqnarray}}
\nc{\eea}{\end{eqnarray}}
\nc{\bal}{\begin{alignedat}}
\nc{\eal}{\end{alignedat}}
\nc{\beq}{\begin{equation}}
\nc{\eeq}{\end{equation}}
\nc{\bit}{\begin{itemize}}
\nc{\eit}{\end{itemize}}
\nc{\benu}{\begin{enumerate}}
\nc{\eenu}{\end{enumerate}}
\nc{\bdes}{\begin{description}}
\nc{\edes}{\end{description}}

\nc{\nn}{\nonumber}

\nc{\hc}{\rm{h.c.}}
\nc{\cc}{\rm{c.c.}}

\nc{\sub}[1]{_{\rm{#1}}}
\nc{\ssub}[1]{_{_\rm{#1}}}
\nc{\super}[1]{^{\rm{#1}}}
\nc{\ssuper}[1]{^{^\rm{#1}}}

\nc{\pare}[1]{\left( #1 \right)}
\nc{\sqpare}[1]{\left[ #1 \right]}
\nc{\ang}[1]{\left\langle #1 \right\rangle}
\nc{\abs}[1]{\left| #1 \right|}

\def\g5{\gamma_{5}}

\def\MeV{\: \rm{MeV}}
\def\GeV{\: \rm{GeV}}
\def\TeV{\: \rm{TeV}}

\def \cm{\: \rm{cm}}

\def \km{\: \rm{km}}
\def \pc{\: \rm{pc}}
\def\kpc{\: \rm{kpc}}

\def \snd{\: \rm{s}}
\def  \yr{\: \rm{yr}}

\def \Gyr{\: \rm{Gyr}}

\def\a{\alpha}

\def\g{\gamma}
\def\d{\delta}

\def\th{\theta}

\def\k{\kappa}
\def\l{\lambda}
\def\m{\mu}
\def\n{\nu}
\def\ks{\xi}

\def\p{\pi}
\def\r{\rho}
\def\s{\sigma}
\def\t{\tau}

\def\x{\chi}

\def \ns{{_\rm{NS}}}
\def \mp{M_\rm{Pl}}
\def \sva{\langle \s v \rangle_a}

\begin{document}


\title{Realistic neutron star constraints on bosonic asymmetric dark matter.}

\author{Nicole F.\ Bell} 
\email{n.bell@unimelb.edu.au}
\affiliation{ARC Centre of Excellence for Particle Physics at the Terascale} 
\affiliation{School of Physics, The University of Melbourne, Victoria 3010, Australia}

\author{Andrew Melatos} 
\email{amelatos@unimelb.edu.au}
\affiliation{School of Physics, The University of Melbourne, Victoria 3010, Australia}

\author{Kalliopi Petraki}
\email{kpetraki@nikhef.nl}
\affiliation{ARC Centre of Excellence for Particle Physics at the Terascale}
\affiliation{School of Physics, The University of Melbourne, Victoria 3010, Australia}
\affiliation{Nikhef, Science Park 105, 1098 XG Amsterdam, The Netherlands}

\preprint{NIKHEF-2013-002}


\begin{abstract}
It has been argued that the existence of old neutron stars excludes the possibility of non-annihilating light bosonic dark matter, such as that arising in asymmetric dark matter scenarios.  If non-annihilating dark matter is captured by neutron stars, the density will eventually become sufficient for black hole formation.  However, the dynamics of collapse is highly sensitive to dark-matter self-interactions. Repulsive self-interactions, even if extremely weak, can prevent black hole formation.  We argue that self-interactions will necessarily be present, and estimate their strength in representative models. We also consider co-annihilation of dark matter with nucleons, which arises naturally in many asymmetric dark matter models, and which again acts to prevent black hole formation. We demonstrate how the excluded region of the dark-matter parameter space shrinks as the strength of such interactions is increased, and conclude that neutron star observations do not exclude most realistic bosonic asymmetric dark matter models.
\end{abstract}



\maketitle

\section{Introduction}
\label{sec:intro}

The similarity of the ordinary matter and dark matter (DM) relic abundances suggests a common origin for both. This motivates the asymmetric dark matter (ADM) scenario, which links dynamically the two matter components of the universe. In this scenario the relic DM abundance is due to a particle-antiparticle asymmetry, in analogy with the relic ordinary matter abundance.  The DM asymmetry is maintained today due to a particle number symmetry governing the low-energy physics of the dark sector, a dark baryon number. The excess of dark particles over antiparticles may be related to the excess of ordinary particles over antiparticles via various mechanisms, thus establishing a tight relation between the dark and ordinary relic densities (see e.g.~\cite{Dodelson:1989cq,*Kuzmin:1996he,*Kitano:2004sv,*Farrar:2005zd,*An:2009vq,
*Gu:2010ft,*Hall:2010jx,*Bell:2011tn,*vonHarling:2012yn,*Petraki:2011mv,
*Heckman:2011sw,*Barr:1990ca,*Barr:1991qn,*Kaplan:1991ah,*Berezhiani:2000gw,
*Foot:2003jt,*Foot:2004pq,*Berezhiani:2008zza,*Cosme:2005sb,*Suematsu:2005zc,
*Gudnason:2006ug,*Gudnason:2006yj,*Banks:2006xr,*Dutta:2006pt,*Khlopov:2008ty,
*Ryttov:2008xe,*Foadi:2008qv,*Kaplan:2009ag,*Cai:2009ia,*Frandsen:2009mi,
*Gu:2010yf,*Dulaney:2010dj,*Haba:2010bm,*Buckley:2010ui,*Chun:2010hz,
*Blennow:2010qp,*Allahverdi:2010rh,*Dutta:2010va,*Falkowski:2011xh,*Haba:2011uz,
*Chun:2011cc,*Kang:2011wb,*Frandsen:2011kt,*Graesser:2011wi,*Graesser:2011vj,*Kaplan:2011yj,
*Cui:2011qe,*Kumar:2011np,*Oliveira:2011gn,*Kane:2011ih,*Barr:2011cz,
*Lewis:2011zb,*Cui:2011wk,*D'Eramo:2011ec,*Kang:2011ny,*Arina:2012aj,*Krylov:2013,
Kusenko:1997si,Davoudiasl:2010am}, and for a review see Ref.~\cite{Davoudiasl:2012uw}). Moreover, ADM models typically predict DM masses of a few GeV. This mass range is currently favoured by the DM direct-detection experiments DAMA, CoGeNT and CRESST~\cite{Savage:2008er,*Aalseth:2010vx,*Angloher:2011uu}.

Observations of stellar objects have been employed to constrain the properties of ADM~\cite{Goldman:1989nd,Kouvaris:2010jy,Kouvaris:2011fi,
Kouvaris:2012dz,McDermott:2011jp,Guver:2012ba,
Sandin:2008db,*Frandsen:2010yj,*Cumberbatch:2010hh,*Taoso:2010tg,*Kouvaris:2011gb,*Zentner:2011wx,
*Iocco:2012wk,*Lopes:2012af,*Casanellas:2012jp}. If DM interacts with nuclear matter, then it can scatter on the nucleons in compact objects, lose energy, and become captured in their interior~\cite{Press:1985ug,*Gould:1987ir,Kouvaris:2007ay}. The DM capture in compact objects may have observable consequences. For example, if DM self-annihilates, its capture and annihilation inside the Sun and the Earth can produce potentially observable neutrino signals (see e.g.~\cite{Cirelli:2005gh,*Blennow:2007tw,*Bell:2011sn,*Bell:2012dk}), while its capture and annihilation in neutron stars can alter the thermal evolution of the latter~\cite{Bertone:2007ae,Kouvaris:2007ay}. 
However, in the ADM scenario, today's universe contains no dark anti-baryons, all of which annihilated with dark baryons at early times. As a result, no DM self-annihilations can take place today.

\emph{Non-annihilating} DM captured in compact objects accumulates over time, without its density being capped by self-annihilations. Depending on the local DM density, the accretion rate of DM in a star may be large enough such that the DM in the core of the star condenses collectively in its ground state. Fermionic matter in its ground state is supported against gravitational collapse by the Pauli exclusion principle, if the total number of particles does not exceed the Chandrashekhar limit $N_\rm{Cha}^f \approx \pare{\mp/m}^3$, where $m$ is the particle mass. Bosonic matter in its ground state is supported by the uncertainty principle, which is overcome if the number of particles exceeds the much lower limit $N_\rm{Cha}^b \approx \pare{\mp/m}^2$. However, this limit is dramatically increased if repulsive self-interactions are present. When the Chandrashekhar limit is exceeded, gravitational collapse ensues, and a black hole can potentially form, unless some other stabilising mechanism takes over. This may be the case if, for example, the particles under consideration are composite rather than fundamental. Then the degeneracy pressure of the constituent degrees of freedoms can eventually withhold gravitational collapse, even when a much higher density is reached.

Following such considerations, Kouvaris and Tinyakov showed that neutron star observations exclude fundamental bosonic ADM having weak-scale interactions with ordinary matter but vanishing self-interactions, in the mass range 2~keV -- 16~GeV~\cite{Kouvaris:2011fi} (see also \cite{McDermott:2011jp}). In this mass range, a black hole would be expected to form in the core of neutron stars located in the Milky Way within their observed lifetime. The black hole would subsequently consume the stars. Importantly, the excluded DM mass range includes the range favoured by DM direct-detection experiments. In the case of self-interacting fundamental bosonic ADM, Ref.~\cite{Kouvaris:2011fi} derived constraints on the DM self-interaction.

\smallskip

In this paper, we point out that:
\benu[(i)]
\item The existence of DM-nucleon interaction, necessary for the capture of DM in stellar objects, implies the existence of DM self-interactions which, if repulsive, are in most cases sufficiently strong to un-exclude the DM mass range of interest. (If the self-interaction is attractive, a different set of considerations is due.)
\item In many ADM models, DM can co-annihilate with nucleons or leptons via the couplings introduced to relate the matter-antimatter asymmetries of the ordinary and the dark sectors in the early universe. The dark-ordinary matter co-annihilations, if sufficiently strong, can cap the DM density in the core of stars and hinder the gravitational collapse.
\eenu

In what follows, we first expound on the above two points: the inevitability of DM self-interactions, and how co-annihilations of DM with ordinary matter arise naturally in the ADM scenario. Following Ref.~\cite{Kouvaris:2011fi}, we then review the DM capture, condensation and gravitational collapse inside the neutron star. We present and comment on the bounds on bosonic ADM in the presence of self-interactions and dark-ordinary matter co-annihilations.

\section{The inevitability of DM self-interaction}
\label{sec:self}

\begin{figure}[t]
\parbox[c]{2cm}{\includegraphics[width=1.8cm]{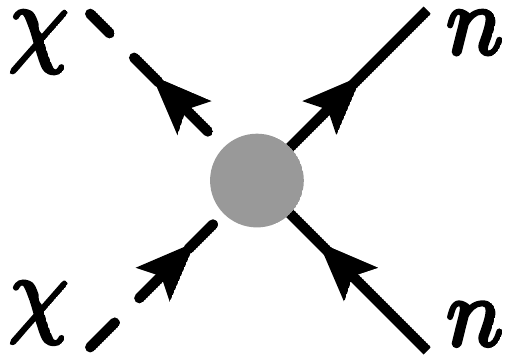}}
\hspace{1cm}
\parbox[c]{3cm}{\includegraphics[width=3.1cm]{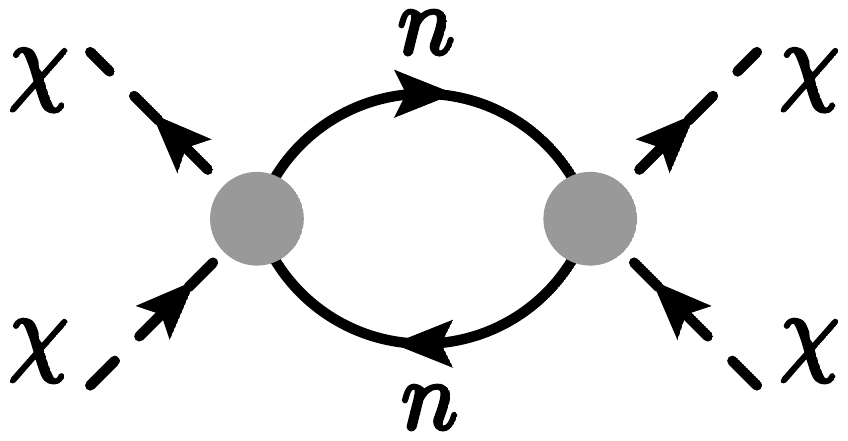}}
\caption{DM-nucleon scattering and DM self-scattering induced by the DM-nucleon effective operator.}
\label{fig:loop}
\end{figure}

If DM scatters elastically with nucleons, $\x n \to \x n$, then it is expected that DM-DM elastic scattering will be mediated by off-shell, as well as on-shell nucleons inside the neutron star, as shown in Fig.~\ref{fig:loop}. However, this contribution to the DM self-interaction is not typically the dominant one. We can obtain a better estimate of the DM self-interaction if we consider various DM-nucleon effective operators and their ultraviolet (UV) completions. 

To put the estimates that will follow in context, it is useful to recall here some relevant figures for the DM-nucleon and DM-DM interactions.\footnote{The figures mentioned in this paragraph refer to a parameter choice leading to rather stringent constraints. In our analysis in the subsequent sections, we make parameter choices that are more robust against astrophysical uncertainties and lead to more conservative constraints.} 
Efficient DM capture in neutron stars takes place for DM-nucleon scattering cross-sections $\s_{n\x} \gtrsim 10^{-45}\cm^2$~\cite{Kouvaris:2007ay}, while no limits apply on bosonic ADM for $\s_{n\x} < 10^{-53}\cm^2$~\cite{Kouvaris:2011fi}. On the other hand, extremely tiny DM-DM interactions can have important effects.  For a repulsive self-interaction of a scalar field $\x$ described by the potential $V_\rm{self} = \l_4 |\x|^4 /4$, with a self-scattering cross-section of $\s_{\x\x} = \l_4^2/(64\p m_\x^2)$, the Chandrashekhar limit is~\cite{Colpi:1986ye}
\begin{equation}
N_\rm{Cha} = \frac{2 \mp^2}{\p m_\x^2} \pare{1 +\frac{\l_4}{32 \p}  \frac{\mp^2}{m_\x^2}}^{1/2}.
\end{equation}
The DM self-interaction affects the bounds on bosonic ADM if the self-scattering cross section is $\s_{\x\x} \gtrsim 10^{-104} \cm^2 (m_\x/\rm{GeV})^2$. In fact, for $\s_{\x\x} \gtrsim 10^{-63}\cm^2$, the DM mass range favoured by direct-detection experiments and predicted in most ADM models, $m_\x \lesssim 10 \GeV$, remains completely unconstrained by neutron stars~\cite{Kouvaris:2011fi}. These values correspond to quartic DM self-couplings $\l_4 \gtrsim 10^{-36}$ and $10^{-16}$ respectively.\footnote{The bounds on $\s_{\x\x}$ are strictly applicable to contact-type DM self-interactions, although they may be considered reasonably applicable also to all short-range interactions, such as e.g. the interaction mediated by a heavy vector boson.} The above indicative figures imply that the DM self-scattering has to be suppressed by many orders of magnitude in respect to the DM-nucleon scattering in order for the former to be insignificant. In this section we argue that this is not the case in most realistic models.

In Table~\ref{tab:interactions}, we list the lowest-order effective operators which can give rise to capture of DM in neutron stars, and the associated DM-nucleon scattering cross-sections~\cite{Goodman:2010ku}. 
For each operator, we consider possible UV completions, and draw the Feynman diagrams which contribute dominantly to the $\x n \to \x n$ and $\x\x\to\x\x$ scattering amplitudes. 

\begin{table*}[t]
\begin{center}
\begin{tabular}{|c|c|c|c|c|c|}
\hline 
\parbox[c]{2.5cm}{ DM-nucleon effective operator }
&\multicolumn{2}{c|}{\parbox[c]{3cm}{\[\frac{\a_s}{4 M_*^2} \, \x^\dagger \x \, G_{\m\n} G^{\m\n}  \] }}
&\multicolumn{2}{c|}{\parbox[c]{3cm}{\[\frac{m_q}{M_*^2} \, \x^\dagger \x \: \bar{q}  q \]}}
&\parbox[c]{3.2cm}{\[\frac{i}{M_*^2} \, \x^\dagger \overleftrightarrow{\partial}_\m \x \: \bar{q} \g^\m  q \]}
\\ 
\hline
\parbox[c]{2.5cm}{ DM-nucleon cross-section, $\s_{n\x}$ }
&\multicolumn{2}{c|}{\parbox[c]{5cm}{\[3 \cdot 10^{-2}  \ \frac{\m_{n\x}^2 m_n^2}{m_\x^2 M_*^4} 
\]}}
&\multicolumn{2}{c|}{\parbox[c]{2.4cm}{\[7 \cdot 10^{-3} \ \frac{\m_{n\x}^2 m_n^2}{m_\x^2 M_*^4}\] }}
&\parbox[c]{2.4cm}{\[\frac{3\m_{n\x}^2}{M_*^4} \]}
\\
\hline
\parbox[c]{2.5cm}{\smallskip possible  \\ UV completions \smallskip}
& \parbox[c]{2.5cm}{scalar  \\ mediation  }
& \parbox[c]{2.5cm}{fermion \\ mediation}
& \parbox[c]{2.5cm}{scalar  \\ mediation}
& \parbox[c]{2.4cm}{fermion \\ mediation}
& \parbox[c]{2.4cm}{vector boson \\ mediation}
\\
\hline
\parbox[c]{2.5cm}{ $\x n \to \x n$ \\ scattering}
&\parbox[c]{2.5cm}{\includegraphics[width=2.4cm]{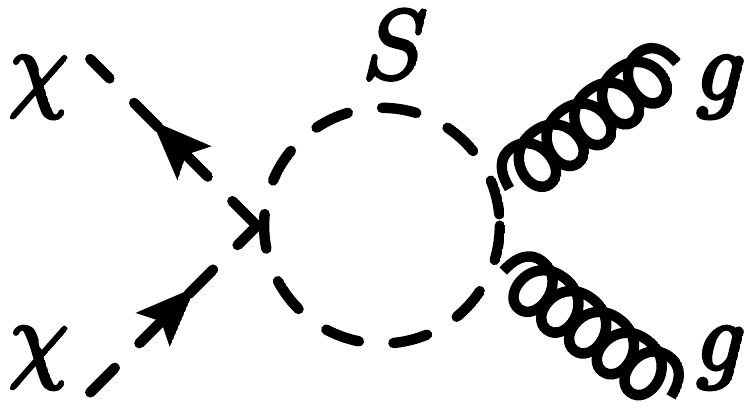} }
&\parbox[c]{2.5cm}{\includegraphics[width=2.1cm]{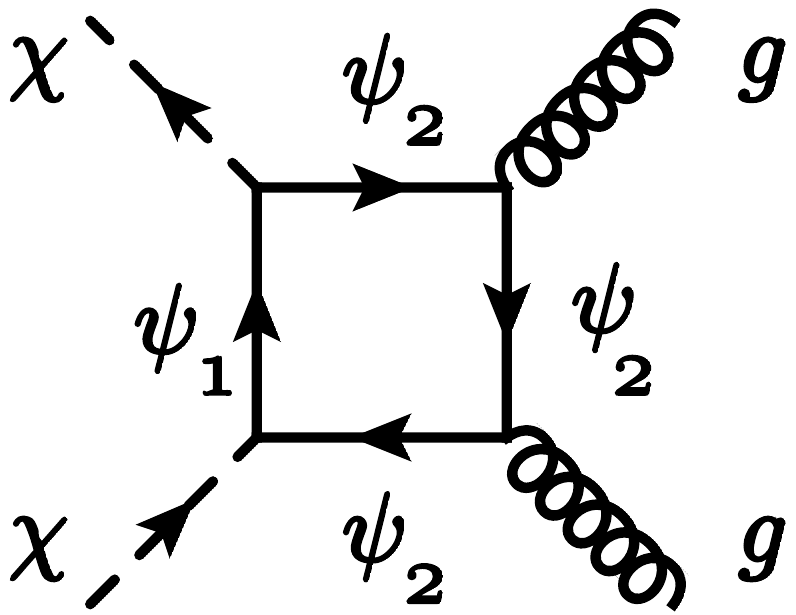}}
&\parbox[c]{2.5cm}{\includegraphics[width=2.5cm]{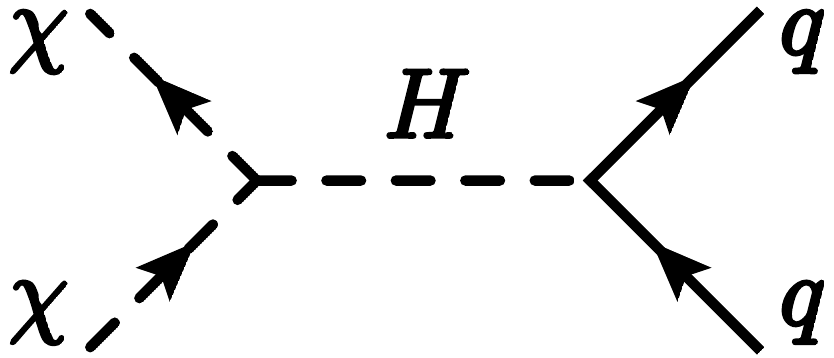}}
&\parbox[c]{2.5cm}{\includegraphics[width=1.5cm]{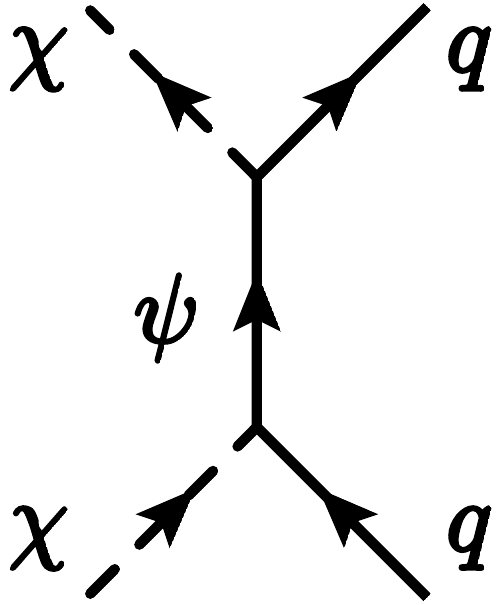}}
&\parbox[c]{2.53cm}{\includegraphics[width=2.5cm]{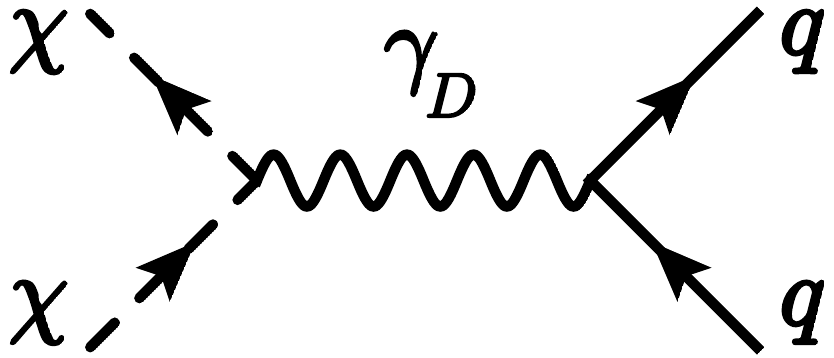}}
\\
\hline
\parbox[c]{2.5cm}{$\x\x \to \x\x$ \\ scattering}
&\parbox[c]{2.5cm}{\includegraphics[width=2.4cm]{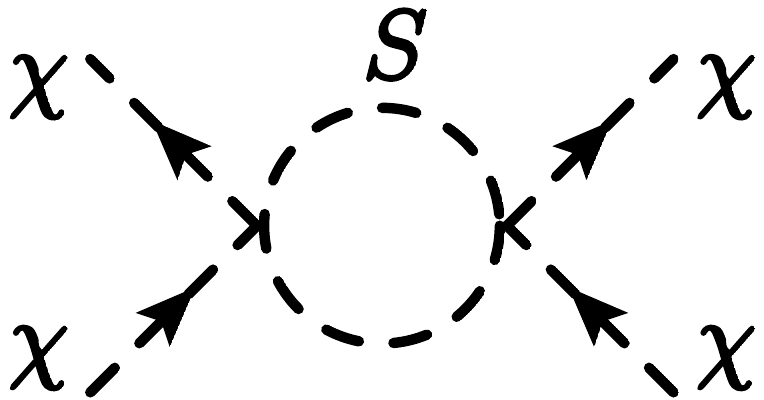}}
&\parbox[c]{2.5cm}{\includegraphics[width=2.1cm]{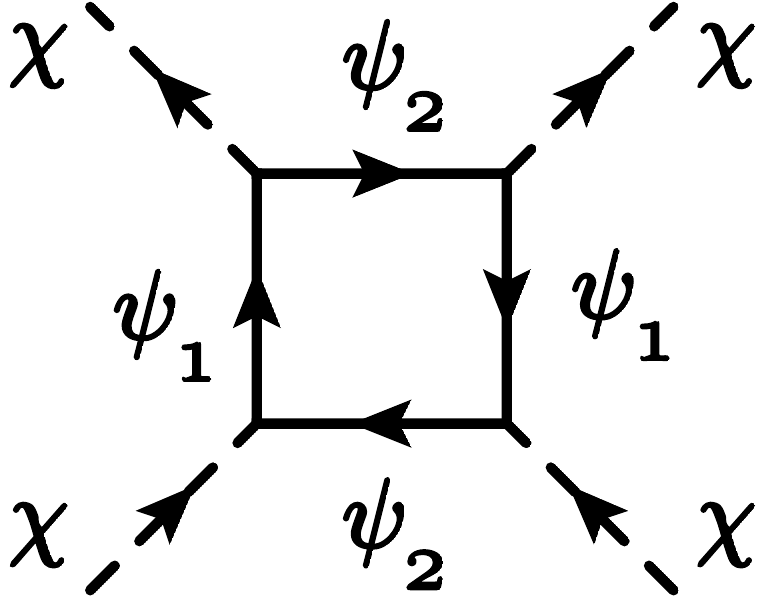}}
&\parbox[c]{2.5cm}{\includegraphics[width=2.5cm]{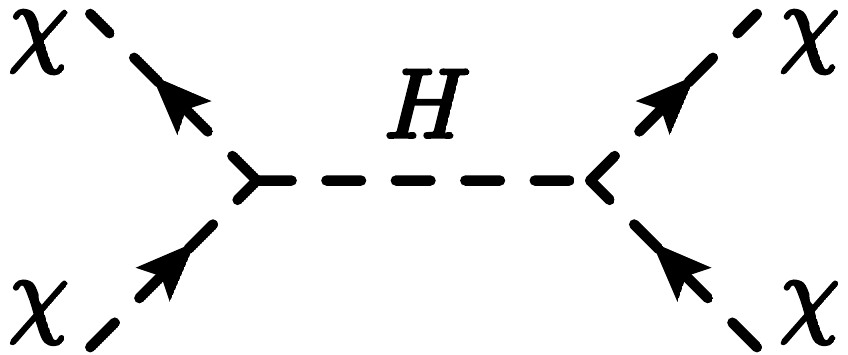}}
&\parbox[c]{2.5cm}{\includegraphics[width=2.1cm]{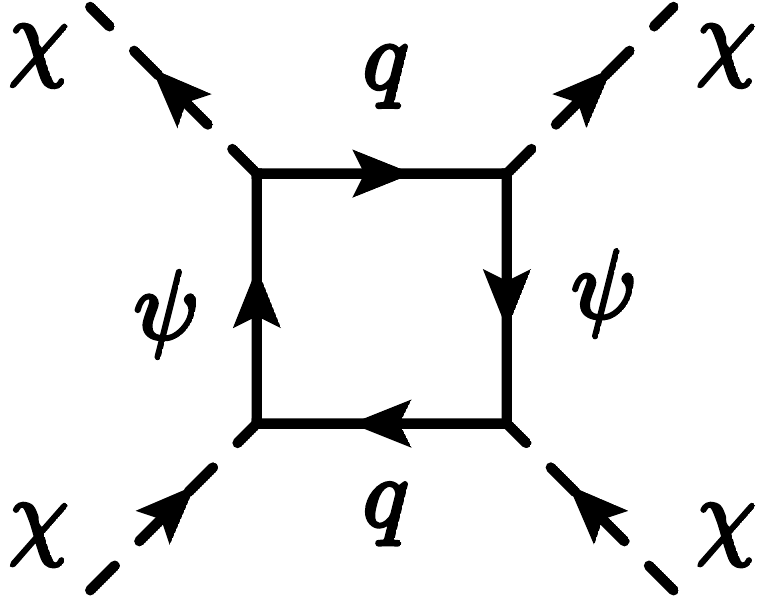}}
&\parbox[c]{2.5cm}{\includegraphics[width=2.5cm]{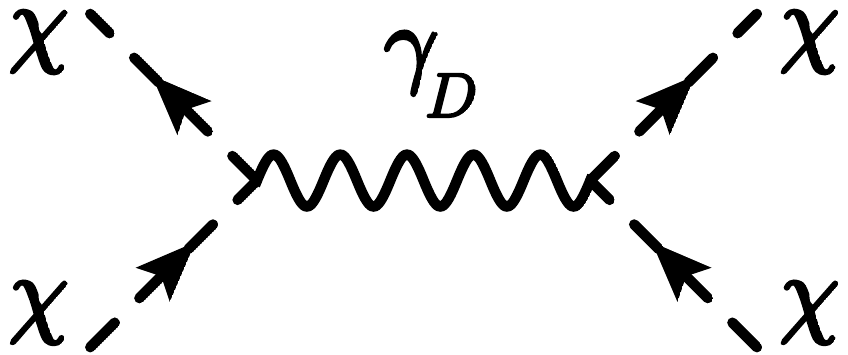}}
\\
\hline
\end{tabular}
\caption{
\label{tab:interactions}
The lowest-order effective interactions of complex scalar DM with nucleons, and the associated DM-nucleon scattering cross-section $\s_{n\x}$, taken from Ref.~\cite{Goodman:2010ku}, where $m_\x, \: m_n$ are the DM and the nucleon mass respectively, and $\m_{n\x}$ is the DM-nucleon reduced mass. For each DM-nucleon operator, we consider possible UV completions and draw the Feynman diagrams which dominantly contribute to DM-nucleon scattering and DM self-scattering. In all cases but one, the $\x n$ and $\x\x$ interactions arise at the same loop order.
(Note that each of the operators presented here has its dual or axial analogue. For those operators, the DM-nucleon cross-section at the non-relativistic limit is velocity-suppressed.)
}
\end{center}
\end{table*}

In all cases but one, DM self-scattering arises at the same loop order as the DM-nucleon scattering. The associated contributions to $\s_{n\x}$ and $\s_{\x\x}$ are then expected to be of comparable magnitude. For example, suppose that the $\x n$ scattering is mediated by a vector boson $Z'$ of mass $m_{_{Z'}}$, with gauge coupling $g_{_{Z'}}$, under which DM and nucleons carry charge $c_\x$ and $c_n$ respectively. The DM-nucleon spin-independent scattering cross-section in the non-relativistic regime is $\s_{n\x} \approx 3 \m_{n\x}^2/M_*^4$, where $\m_{n\x}$ is the nucleon-DM reduced mass, $M_*^2 = m_{_{Z'}}^2 / g_{_{Z'}}^2 c_\x c_n$, and the numerical factor is determined after the nuclear form factors have been appropriately taken into account~\cite{Goodman:2010ku}. The DM self-scattering cross-section is
\bea
\s_{\x\x} 
&\simeq & \frac{c_\x^4 g_{_{Z'}}^4 m_\x^2}{\p m_{_{Z'}}^4} \nn \\ 
&\approx&  10^{-46} \cm^2 \pare{\frac{\s_{n\x}}{10^{-45} \cm^2}}
\pare{\frac{m_\x^2}{\m_{n\x}^2}}
\pare{\frac{c_\x^2}{c_n^2}}  \:  ,
\eea
which, in view of the characteristic values for the DM self-interaction mentioned above, implies that bosonic ADM having vector interactions with nucleons is rather unconstrained by neutron star observations. 
Even if $\x\x$ scattering arises at one extra loop order in comparison to $\x n$ scattering, as in the case of scalar or pseudo-scalar DM-quark coupling mediated by a heavy fermion\footnote{We note that this particular DM self-interaction is attractive, and this case is in fact not subject to the analysis and the constraints of Refs.~\cite{Kouvaris:2011fi,McDermott:2011jp}.}, 
the one-loop suppression is unlikely to account for the many orders of magnitude suppression needed to render the DM self-interaction unimportant.

The above suggest that the DM self-interaction induced by the couplings responsible for the DM-nucleon scattering, is likely sufficiently strong to render light bosonic ADM viable. This appears to be true unless some symmetry forbids the DM self-interaction or cancels the above contribution. The DM self-interaction arising from couplings responsible for the $\x n$ scattering, generates the operator $|\x|^4$.
For a fundamental boson $\x$, it is not possible to forbid the renormalisable self-interaction term $|\x|^4$ by requiring invariance under any unitary symmetry transformation of the field. This is in fact why a proper calculation including loop corrections shows that the induced quartic coupling always diverges. 
The $|\x|^4$ operator could possibly be eliminated only by a space-time symmetry, such as supersymmetry (SUSY). We shall now discuss both cases.\footnote{In the case of vector DM-quark interaction, the non-renormalisable DM effective operator $(\x^\dagger \overleftrightarrow{\partial} \x)^2$ is also generated at tree level (and could potentially give the dominant contribution to the DM self-scattering).}

\medskip

\paragraph{Non-supersymmetric theories: }

No symmetry disallows the $|\x|^4$ operator. For a field possessing no interactions with other fields, the quartic self-coupling can be consistently set to zero, as it will not be generated by any interactions. However, for a field interacting with other particles -- as has to be the case for DM which can scatter off nucleons -- a $|\x|^4$ term is inevitably generated. Theoretical consistency then requires that the $|\x|^4$ term is included in the bare Lagrangian, together with the corresponding counterterm which, after renormalisation, cancels any divergences present and sets the self-coupling to its physical value. 

The physical value of the $|\x|^4$ coupling, being renormalisable, remains of course a free parameter, and as such it could be vanishingly small. However, the detailed cancellation of the various contributions to the $|\x|^4$ coupling amounts to fine tuning. Even if set to zero at a particular energy, the value of the $|\x|^4$ coupling will in fact run to non-vanishing values at different energies. How this running affects the fate of the DM condensate inside the neutron star\footnote{The formation of a DM Bose-Einstein condensate is necessary for gravitational collapse (see Refs.~\cite{Kouvaris:2011fi,Kouvaris:2012dz} and sections \ref{sec:CaptureThermCond} and \ref{sec:collapse}).\label{foot:cond}} 
can be seen best in the field-space description, to which we now turn.

Consider, for example, the DM-gluon effective interaction of Table~\ref{tab:interactions}, $(\a_s/4M_*^2) \, \x^\dagger \x \, G_{\m\n}G^{\m\n}$, mediated by a heavy SU(3)$_c$ charged scalar field $S$ of mass $m_S$, which couples to the DM field by 
\beq
V_{_\rm{UV}} \supset \k |\x|^2 |S|^2 \ .
\label{eq:UV}
\eeq 
Then $1/M_*^2 = (11 c/6\p)\,\k/m_S^2$, where $c=4/3$ if $S$ transforms as a triplet under SU(3)$_c$.\footnote{Note that for the DM-gluon coupling mediated by a scalar, there are two more diagrams contributing to the $\x n$ scattering which are not shown in Table~\ref{tab:interactions}: one with the gluons crossed, and one with the two gluons attached to the same vertex.} We may calculate the 1-loop effective potential $V_\rm{eff} (\x)$ according to standard methods~\cite{Coleman:1973jx}. 
We first express the complex field $\x$ in terms of amplitude and phase $\x = (\tilde{\x}/\sqrt{2}) e^{i \th}$. We shall include a counterterm for the quartic interaction, $C_4 |\x|^4$, and impose renormalisation conditions 
\beq
\left. \frac{d^2 V_\rm{eff}}{d\tilde{\x}^2}\right|_{\tilde{\x} = 0} = m_\x^2 \ , \quad
\left. \frac{d^4 V_\rm{eff}}{d\tilde{\x}^4}\right|_{\tilde{\x} = \x_0} = 0 \ .
\eeq
The renormalisation conditions mean that for $\ang{\x} = \x_0/\sqrt{2}$ the quartic self-interaction of the $\x$ field vanishes. (Of course, $\x_0$ can be chosen to be 0.)
Under these conditions, the effective scalar potential becomes
\bea
V_\rm{eff} &=& m_\x^2 |\x|^2 + \frac{ m_S^4}{64 \p^2} \left\{\pare{1 + 2 w}  \ln \pare{1+w} -  w \right. \nn \\ 
&+& \left.w^2 \sqpare{\ln \pare{\frac{1 + w}{1+w_0}} - \frac{25w_0^2+42w_0+9}{6 \pare{1+w_0}^2}}  \right\}
\label{eq:Veff}  
\eea
where we have defined 
\beq
w   \equiv \frac{\k\x^2}{m_S^2}  
\quad \rm{and} \quad
w_0 \equiv \frac{\k \pare{\x_0/\sqrt{2}}^2}{m_S^2}  
\ . 
\label{eq:w,w0} 
\eeq

For definiteness, let us choose $\x_0 = 0$. 
At field values $\x \ne 0$, the quartic self-coupling, $\l_4  \equiv  \pare{d^4 V_\rm{eff} / d\tilde{\x}^4} / 3$, is  
\beq
\l_4 \simeq \frac{\k^2}{12\p^2} \sqpare{ \frac{w(w+3/2)}{(1+w)^2} +\frac{3}{8}\ln (1+w)}
\  .
\label{eq:l4 non-susy}
\eeq
The parameter $w$ in \eq{eq:l4 non-susy} is related to the DM-nucleon cross-section, $\s_{n\x}$, and the expectation value of the field $\x$, which is  the order parameter of the condensation
\beq
w
\approx \pare{\frac{\ang{\x}}{7 \TeV}}^2 
\pare{\frac{\s_{n\x}}{10^{-45} \cm^2}}^{1/2}
\pare{\frac{m_\x}{\m_{n\x}}}
\ ,
\label{eq:w}
\eeq
where we used  $\s_{n\x} \approx 3\cdot 10^{-3} \m_{n\x}^2 m_n^2 /(m_\x^2 M_*^4)$~\cite{Goodman:2010ku}, or $\s_{n\x} \approx 2 \cdot 10^{-3} \k^2  \m_{n\x}^2 m_n^2 /(m_\x^2 m_S^4)$.
Note that at $w > 1$, $\l_4 \sim \k^2/12 \p^2$. 

We shall now estimate $\ang{\x}$. Assuming vanishing self-interactions, the (low-energy) Lagrangian density of the DM field is $\cal{L} = \partial^\m \x^* \partial_\m \x - m_\x^2 |\x|^2$. Ignoring finite-volume effects, which become important only when the condensate becomes relativistic, the condensate is described by a spatially-homogeneous  field configuration $\x = (\tilde{\x}/\sqrt{2}) \exp (i \theta)$. The number of particles in the condensate is Noether's conserved charge, $N_\rm{cond} = i \int d^3x \sqpare{\pare{\partial^0 \x^*}\x - \x^* \pare{\partial^0 \x}} = \tilde{\x}^2 \dot{\theta}\:\cal{V}$, where $\cal{V}$ is the volume of the condensate. The Klein-Gordon equation yields $\dot{\theta}^2 = m_\x^2$, thus\footnote{For the spatially homogeneous field configuration, the kinetic and the potential energy densities are $\cal{E}_k =\partial_\m \x^* \partial^\m \x = (\tilde{\x}^2/2) \dot{\th}^2$ and $\cal{E}_p = m_\x^2 \tilde{\x}^2/2$ respectively, which implies $\cal{E}_k = \cal{E}_p$, as expected from the virial theorem for a harmonic potential. Then \eq{eq:cond energy} can be re-expressed as $\cal{E}_\rm{tot} = \cal{E}_k + \cal{E}_p = \pare{N_\rm{cond} / \cal{V}} m_\x$. This describes a collection of zero-momentum particles, as expected.}
\beq 
\ang{\x}^2 \simeq N_\rm{cond}/2 m_\x \cal{V}~. 
\label{eq:cond energy}
\eeq 
Obviously, $\ang{\x}$ increases as more DM particles accumulate in the ground state. 
Eventually, when close to gravitational collapse, $N_\rm{cond} \approx N_\rm{Cha} \approx (\mp/m_\x)^2$, and the size of the condensate is $r_c \sim 1/m_\x$~\cite{Colpi:1986ye}, assuming again vanishing self-interaction.\footnote{In the early stages of condensation, when the neutron star gravity still dominates over the gravity of the condensate, the size of the condensate can be estimated from the spread of the wavefunction of the DM ground state in the harmonic gravitational potential in the interior of the star, $r_0 \approx (8\p G \r_\ns m_\x^2/3)^{-1/4}$~\cite{Kouvaris:2011fi}. In this stage, $1/r_0 \sim 10^{-10} \GeV (m_\x/\rm{GeV})^{1/2} \ll m_\x$, for all of the mass range of interest, and the zero-momentum approximation for the condensate is valid. When close to the Chandrashekhar limit, the gravity of the condensate has become larger than the neutron star gravity, and the size of the condensate is $r_c \sim 1/m_\x$. The condensed particles become quasi-relativistic, and only then the homogeneous approximation starts breaking down. See sections~\ref{sec:condensation} and \ref{sec:collapse} for more details.}
At this stage, $\ang{\x} \sim 10^{-2}\mp$~\cite{Colpi:1986ye}. Thus, as condensation proceeds, the parameter $w$ takes values $w \gg 1$.

The above implies that a significant quartic self-coupling is generated as the number of DM particles in the condensed state increases
\beq
\l_4 \sim 8 \cdot 10^{-3} \, \k^2  \  .
\label{eq:l4 non-susy num}
\eeq
The coupling $\k$ cannot be arbitrarily small in order for $\s_{n\x}$ to be sizable. The rough experimental lower limit on new coloured particles, $m_S > 1\TeV$, implies $\k^2 \gtrsim 10^{-3} \pare{\s_{n\x} / 10^{-45} \cm^2} \pare{m_\x^2 / \m_{n\x}^2}$.
Thus, the estimate of \eq{eq:l4 non-susy num} suggests that $\l_4$ takes values many orders of magnitude greater than is necessary to prevent gravitational collapse of light bosonic ADM in neutron stars. Note that we would have obtained a similar estimate for $\l_4$ generated by the variation of the $\x$ expectation value, had we chosen the field value at which $\l_4$ is set to vanish to be $\x_0 \ne 0$.

There is an alternative description of the same concept: The running of the quartic coupling is due to the higher-order effective operators, $|\x|^n$ with $n>4$, generated by the renormalisable couplings of DM to itself and to other fields. The values of the non-renormalisable couplings are determined once the physical value of the renormalisable $|\x|^4$ coupling is set. Similarly to the quartic self-interaction, higher-order self-couplings also alter the Chandrashekhar limit for gravitational collapse of bosonic particles, and can thus prevent the black-hole formation inside neutron stars~\cite{Ho:1999hs} (see \eq{eq:M Cha}). For the effective potential of \eq{eq:Veff}, we may estimate the coupling $\l_6  \equiv (2^3/6!) \pare{d^6 V_\rm{eff} / d\tilde{\x}^6}$ at $\tilde{\x}= \x_0 = 0$
\beq
\l_6 \sim \frac{\k^3}{32 \p^2 \, m_S^2}  \ .
\label{eq:l6 non-susy}
\eeq  
The magnitude of this coupling can be more easily compared with the bounds presented in Figs.~\ref{fig:NuChi}-\ref{fig:Coann} below, when expressed in terms of the dimensionless combination $\l_{\x\x}=32\p \pare{\l_6 m_\x^2/48\p}^{1/2}$ (see \eq{eq:self coupling}), to which the same constraints as for $\l_4$ apply:
\beq
\l_{\x\x} \! \sim 10^{-3} \, \k
\!\pare{\frac{\s_{n\x}}{10^{-45}\cm^2}}^\frac{1}{4}
\!\! \pare{\frac{m_\x}{\m_{n\x}}}
\!\! \pare{\frac{m_\x}{\rm{GeV}}}  
\ .
\label{eq:l6 non-susy num}
\eeq

\smallskip

The above analysis can of course be repeated for any interaction giving rise to DM-nucleon scattering, and will lead to similar conclusions.

\medskip

\paragraph{Supersymmetric theories: }

Supersymmetry can assert, under certain conditions, the absence of scalar self-couplings, even if the scalar field possesses interactions with other particles. 
The directions of the scalar potential along which quartic and possibly higher-order operators are absent are known as ``flat directions''. Flat directions (FDs) are quite common in SUSY models, with the minimal supersymmetric standard model having a large number of them~\cite{Gherghetta:1995dv}. The absence of self-coupling along FDs is typically due to the combined effect of gauge symmetries and supersymmetry. Moreover, due to non-renormalisation theorems governing supersymmetric theories, it is possible to set even gauge-invariant supersymmetric couplings to zero without them being generated by other interactions~\cite{Salam:1974jj,*Grisaru:1979wc}.

In either case, the absence of self-couplings along FDs in SUSY theories should not be interpreted as an absence of contributions to these couplings induced by the interactions of the FD fields with other particles. It can rather be understood as cancellation between diagrams contributing to the same coupling. Cancellations can occur between diagrams which are related by a SUSY transformation, i.e. diagrams in which the degrees of freedom propagating off-shell belong to the same supermultiplet. Exact cancellation is warranted by exact supersymmetry.

However, supersymmetry -- if at all present in nature -- is broken. This is evident for the sector of ordinary particles, described by the standard model (SM). The interactions of DM with SM particles ensure that SUSY breaking is transmitted to the dark sector as well (even if this is not happening via more direct couplings of the dark sector to the hidden sector responsible for the SUSY breaking). In fact, SUSY has to be broken in the dark sector in order for DM to be purely bosonic. Supersymmetry implies that for every boson there is a fermion which carries the same additive quantum numbers, and has the same mass. SUSY breaking lifts the degeneracy between the masses of the bosonic and the fermionic components of a supermultiplet. In the scenario we are considering, this renders the bosonic component of the DM supermultiplet the lightest degree of freedom carrying dark baryon number, into which all other dark baryons decay and impart their dark baryonic charge.

The quartic couplings induced by the SUSY-breaking hidden sector are typically
\beq
\l_4\sim 
\left\{
\bal{5}
&\frac{m_s^2}{\mp^2}&  \: = \:  &10^{-32} \pare{\frac{m_s}{\rm{TeV}}}^2&   , \quad  &\rm{PMSB}& \\
&\frac{m_s^2}{M_m^2}&  \: = \:  &10^{-16} 
\sqpare{\frac{m_s/\rm{TeV}}{M_m / (10^{11} \GeV)}}^2  
&   , \quad &\rm{GMSB}& \ , \\
\eal
\right.
\label{eq:lambda susy}
\eeq
in the Planck-scale mediated SUSY breaking (PMSB) and gauge-mediated SUSY breaking (GMSB) scenarios~\cite{deGouvea:1997tn,*Enqvist:1997si}. 
Here $m_s$ is the soft scale of SUSY-breaking and $M_m$ is the scale of the messenger fields in the GMSB case.\footnote{Flat directions are expected to be lifted also by supersymmetric non-renormalisable effective operators, which would contribute to the self-interaction of the FD field. However, the existence of such operators depends on assumptions about the physics beyond the realm of a certain model, and we shall not invoke them further in our discussion.} 

\smallskip

The estimated couplings of \eqs{eq:l4 non-susy num}, \eqref{eq:l6 non-susy num} and \eqref{eq:lambda susy} can lift the exclusion from a significant portion of the parameter space~\cite{Kouvaris:2011fi}. We shall discuss this in more detail in the subsequent sections.

\section{DM co-annihilations with nucleons or leptons, in the ADM scenario}
\label{sec:co}

Establishing a tight connection between the ordinary and the dark matter-antimatter asymmetries -- the main motivation of the ADM scenario -- requires particular symmetry structures and symmetry breaking patterns. Here we discuss how these patterns could imply that DM can co-annihilate with nucleons or leptons today. 

The excess of particles over antiparticles in each sector is sustained due to a global continuous abelian symmetry: the ordinary baryon number symmetry $B_v$ of the visible sector, and, analogously, the dark baryon number symmetry $B_d$ of the dark sector. In order for $B_v$ and $B_d$ net charges to be generated, these symmetries must be violated by some interactions which took place in the high-energy environment of the early universe, but are ineffective in today's low-energy universe. For the $B_v$ and $B_d$ net charges of our universe to be related, the $B_v$ and $B_d$ violation should not be independent. That is, there should be high-energy interactions which violate both $B_v$ and $B_d$, but preserve a linear combination of the two. 

$B_v$ has a particularity: despite being respected by all perturbative processes of the SM,  it is anomalous and is violated by non-perturbative effects called sphalerons. Sphalerons violate also the lepton number of the SM, but preserve the combination $(B-L)_v$. Cosmologically, they were operative only before the electroweak phase transition of the universe. This means that if a net $B_v$ or $L_v$ charge was generated at an earlier time, it must have been partially converted into a net $L_v$ or $B_v$ charge respectively. Because of sphalerons, it makes sense to replace in our discussion the anomalous $B_v$  with the the anomaly-free $(B-L)_v$.

Processes which violate $(B-L)_v$ and $B_d$ but preserve a linear combination of the two, should generate, after the heavier degrees of freedom have been integrated out, effective operators of the form $\cal{L}_a = \cal{O}_{q_v} (\text{SM}) \, \cal{O}_{q_d} (\text{DS})$. Here 
$\cal{O}_{q_v} (\text{SM})$ is an operator involving SM fields that is invariant under the SM gauge group and carries charge $q_v$ under $(B-L)_v$. Possibilities include the fermionic operators 
$
\overline{u^c}\!\!_R^{} d_R^{} s_R^{}	\,  ,  \:
\bar{\ell}_L^{} \ell_L^c  e_R^{}	\,  ,  \:
\bar{\ell}_L^{} Q_L^{c}  d_R^{}		\,  ,  \:
H \bar{\ell}_L				\,  ,
$
all of which carry $q_v=1$, or some product of them.  $\cal{O}_{q_d} (\text{DS})$ is an operator involving dark-sector fields, among them the DM field $\x$, that is invariant under the dark-sector gauge group and carries charge $q_d$ under $B_d$.

The effective operator $\cal{L}_a$ can induce co-annihilations of DM with either SM baryons or leptons. 
For clarity, let us consider $B_d$ to be normalised such that the DM field $\x$ carries charge unity under it. If $q_v,q_d = \pm 1$, then one DM particle can co-annihilate with one SM baryon or lepton into radiation, i.e. light relativistic particles of both sectors which do not carry $(B-L)_v$ or $B_d$. This occurs in various ADM models.
In the model of Refs.~\cite{Davoudiasl:2010am,Davoudiasl:2011fj}, the operator inducing DM-nucleon co-annihilation (referred to as ``induced nucleon decay'') is also responsible for the simultaneous generation and correlation of the visible and dark baryonic asymmetries in the early universe. Efficient visible and dark baryogenesis suggests a range of values for the strength of this operator, albeit this range spans many orders of magnitude. It follows that the possible range for the co-annihilation cross-section also spans many orders of magnitude. As we will show, for a large range of suggested co-annihilation cross-sections, the DM-nucleon or DM-lepton co-annihilations inside neutron stars can significantly affect the DM concentration, and alter current bounds.

If $q_v, q_d \neq \pm 1$, $\cal{L}_a$ preserves a discrete subgroup of $(B-L)_v$ and/or $B_d$. Co-annihilations would then have to involve more than one nucleon/lepton and/or more than one DM particle. This of course greatly suppresses their rate. We do not consider this possibility further.

\section{DM capture, thermalisation and condensation inside the neutron star}
\label{sec:CaptureThermCond}

\subsection{Capture}
\label{sec:capture}

The accretion of DM particles in the neutron star is described by the equation
\beq
\frac{d N_\x}{dt} = C_{n\x} + C_{\x\x} N_\x -C_a N_\x \ ,
\label{eq:capture}
\eeq
where $N_\x$ is the number of DM particles captured in the star. $C_{n\x}$ is the capture rate due to the scattering of DM on nucleons, $C_{\x\x}$ is 
due to incident DM scattering on DM particles already captured in the neutron star (self-capture), 
and $C_a$ is the DM-nucleon or DM-lepton co-annihilation rate per DM particle.

\setcounter{paragraph}{0}
\paragraph{Capture on nucleons:}
$C_{n\x}$ is estimated to be~\cite{Kouvaris:2007ay} 
\beq
C_{n\x}  = \sqrt{6 \p} \pare{\frac{\r_\x}{m_\x} } \frac{1}{\bar{v}_\x} 
\: \pare{ \frac{2 G M_\ns R_\ns}{1-\frac{2G M_\ns}{R_\ns}}} 
\, f \ ,
\label{eq:C nx}
\eeq
where $\r_\x, \bar{v}_\x$ are the DM density and average velocity in the vicinity of the star, and $m_\x$ is the DM mass. $M_\ns, \ R_\ns$ are the mass and radius of the neutron star. The efficiency factor $f$ takes into account the saturation of the capture rate at sufficiently large cross-sections, when all incident DM is captured
\beq
f = \left\{
\bal{5}
1 				\, , & \quad \s_{n\x} > \s_\rm{sat} \equiv  \frac{R_\ns^2}{0.45 N_{_\rm{B}} \, \ks} \approx \ks^{-1} \, 10^{-45} \cm^2  \\
\frac{\s_{n\x}}{\s_\rm{sat}} 	\, , & \quad \s_{n\x} < \s_\rm{sat}  
\eal
\right.
\label{eq:fn}
\eeq
with $N_{_\rm{B}} \approx 10^{57}$ the number of baryons in the neutron star. The factor $\ks$ represents the suppression in the capture rate due to the neutron degeneracy: if the momentum transfer $\d p$ between nucleons and DM is low, only neutrons with momentum $p > p_{_\rm{F}} - \d p$ participate in the capture process. Here, $p_{_\rm{F}} \simeq (3 \p^2 n_{_\rm{B}})^{1/3} \approx 0.5 \GeV$ is the Fermi momentum, and 
$\d p \approx \sqrt{2} \m v_\rm{esc} \sim \m$, where $\m$ is the DM-nucleon reduced mass and $v_\rm{esc} = (2G M_\ns/R_\ns)^{1/2} \approx 0.6$ is the escape velocity from the surface of the neutron star, which approximates the velocity with which DM particles reach the neutron star. Thus
\beq
\ks 
\simeq   \min \sqpare{1- \pare{1- \frac{\d p}{p_{_\rm{F}}}}^3, 1}   
\approx  \min \sqpare{\frac{m_\x}{0.2 \GeV}, 1} \ . 
\label{eq:ksi}
\eeq

\paragraph{Self-capture:} 
$C_{\x\x}$ is the capture rate due to incident DM particles scattering on DM particles already captured in the neutron star. $C_{\x\x}$ has been estimated in Ref.~\cite{Zentner:2009is}. It becomes important only if $C_{\x\x} \t_\ns \gtrsim 1$, where $\t_\ns$ is the neutron star lifetime. This requires DM self-scattering cross-section 
$
\s_{\x\x}~\!\!\gtrsim~\!\!10^{-34}~\rm{cm}^2 
\pare{	\frac{m_\x}{\rm{GeV}}			}
\pare{	\frac{ 100 \GeV/\cm^3 }{\r_\x} 	}
\pare{	\frac{\bar{v}_\x}{100 \km/\snd}		}
\pare{	\frac{\rm{Gyr}}{\t_\ns}			} 
$ (see e.g.~\cite{Guver:2012ba}), where $\r_\x, \bar{v}_\x$ are the DM density and average velocity in the vicinity of the star. At such high values of $\s_{\x\x}$, the bounds from potential gravitational collapse of DM inside a neutron star are offset far beyond the mass range of interest~\cite{Kouvaris:2011fi}. In the following, we shall thus ignore the DM self-capture entirely.

\paragraph{Co-annihilation rate:}
$C_a$ is the DM-nucleon or DM-lepton co-annihilation rate per DM particle,
\beq
C_a   = \sva \: n_{_\rm{B,L}}  \label{eq:C a} \ ,
\eeq
where $\sva$ is the DM-nucleon or DM-lepton co-annihilation cross-section times relative velocity, averaged over the (thermal) distribution of DM and nucleons or leptons in the neutron star. Here $n_{_\rm{B}} \sim 10^{38} \cm^{-3}$ and $n_{_\rm{L}} \sim 10^{36} \cm^{-3}$ are the baryon and lepton number densities in the core of the neutron star. In the following, for definiteness we will consider DM co-annihilation with nucleons. For DM co-annihilation with leptons, the values of $\sva$ in the bounds of Figs.~\ref{fig:NuChi}-\ref{fig:Coann} should be rescaled by a factor of about $10^2$.

Ignoring self-capture, \eq{eq:capture} yields

\beq
N_\x (t) = \frac{C_{n\x}}{C_a } \sqpare{1-e^{-C_a  t}}  \  .
\label{eq:Nchi}
\eeq
Thus, the total DM mass captured in the neutron star is roughly (we use the exact expression for numerical calculations)
\begin{widetext}
\begin{multline}
M_\rm{capt} \approx 7 \cdot 10^{45} \GeV
\pare{\frac{\r_\x}{100 \GeV/\cm^3 }}
\pare{\frac{100 \, \km/\snd }{\bar{v}_\x}} \times
\\
\min \sqpare{1 \, , \, \frac{\s_{n\x}}{10^{-45} \cm^2} \, , \, \frac{\s_{n\x}}{10^{-45} \cm^2} \, \frac{m_\x}{0.2 \GeV}}
\min \sqpare{\frac{\t_\ns}{10\Gyr} \, , \, \frac{10^{-56} \rm{cm}^3/\rm{s}}{\sva}}   \  .
\label{eq:Mcapt num}
\end{multline}
\end{widetext}

\subsection{Thermalisation}
\label{sec:thermal}

The DM particles captured in the neutron star thermalise via their collisions with neutrons. We may estimate the thermalisation time as follows. The momentum loss per collision is $\d p/\d \n \sim - \sqrt{2}\m_{n\x} v$, where $\n$ is the number of collisions. The rate of collisions is $d\n/dt \simeq \s_{n\x} v \, \tilde{n}_{_\rm{B}}$, where $\tilde{n}_{_\rm{B}}$ is the number of neutrons available to absorb energy. As the DM particles slow down, the momentum transfer drops below the Fermi momentum, and $\tilde{n}_{_\rm{B}}$ is then only a fraction of all neutrons, $\tilde{n}_{_\rm{B}} \sim \tilde{\ks} n_{_\rm{B}}$, where $\tilde{\ks} \sim 3\sqrt{2} \m_{n\x} v/p_{_\rm{F}}$. The total thermalisation time is dominated by the late stages of thermalisation. Putting everything together we find that the DM particles reach thermal velocities $v_\rm{th} = (3 T_\ns /m_\x)^{1/2}$ in 
\bea
\t_\rm{th} 
&\sim& \frac{m_\x^2 p_{_\rm{F}}}{9 \m_{n\x}^2 \s_{n\x} n_B T_\ns} \nn \\
&\sim& 20 \snd 
\pare{\frac{m_\x^2}{\m_{n\x}^2}} 
\! \pare{\frac{10^{-45}\cm^2}{\s_{n\x}}}
\! \pare{\frac{10^7 \: \rm{K}}{T_\ns}} \ ,
\label{eq:t thermal}
\eea
where $T_\ns$ is the temperature in the core of the neutron star.
Thermalisation occurs if $\t_\rm{th} \lesssim \min (\t_\ns, \ C_a^{-1})$, i.e. for 
\begin{multline}
\s_{n\x} \gtrsim 10^{-61} \cm^2 
\pare{\frac{m_\x^2}{\m_{n\x}^2}}
\pare{\frac{10^7 \: \rm{K}}{T_\ns}} \times  \\
\max \sqpare{\frac{10 \Gyr}{\t_\ns}, \: \frac{\ang{\s v}_a}{10^{-56}\cm^3/\snd}}  
\ .
\label{eq:therm condition}
\end{multline}

The thermalised DM is concentrated within a radius $r_\rm{th}$, which can be estimated from the virial theorem in the harmonic gravitational potential in the interior of the star, $GM_\ns m_\x r_\rm{th}^2 / 2 R_\ns^3 = m_\x v_\rm{th}^2 / 2 = 3T_\ns/2$, thus
\beq
r_\rm{th} 
\simeq 30 \: \rm{m} \pare{\frac{T_\ns}{10^7 \: \rm{K}}}^{1/2}  \pare{\frac{\rm{GeV}}{m_\x}}^{1/2}   .
\label{eq:rth}
\eeq
For small $m_\x$, $r_\rm{th}$ exceeds the radius of the star and DM evaporates. We will consider constraints only for $r_\rm{th} \lesssim  0.2 R_\ns$, or
\beq 
m_\x \gtrsim 0.3 \MeV \pare{\frac{T_\ns}{10^7 \: \rm{K}}}
\ . 
\label{eq:thermal evap} 
\eeq

\subsection{Condensation}
\label{sec:condensation}

As DM particles accumulate in the interior of the star, their density increases, and it is possible that at some point it exceeds the critical density for condensation. For vanishing self-interactions, this is 
$n_{_\rm{BEC}} = 2.6 \, (m_\x T_\ns/2\p)^{3/2}$. Then the number of particles needed for the formation of a Bose-Einstein condensate (BEC) in the interior of the star is $N_{_\rm{BEC}} = (4\p/3) r_\rm{th}^3 \, n_{_\rm{BEC}} \simeq 3 \cdot 10^{42} \pare{T_\ns/10^7 \: \rm{K}}^3$.
All DM particles captured in the star in excess of this number are added to the condensed state, as long as the BEC is dilute and depletion is negligible.\footnote{The condition for diluteness is $3 a^3 N_\rm{cond} / (4 \p r_c^3) \ll 1$. Here $a = (\sigma_{\chi \chi}/ 4 \pi)^{1/2}$ is the $s$-wave scattering length. For quartic self-interaction, $\s_{\x\x} = \l_4^2/64\p m_\x^2$. $N_\rm{cond}$ and $r_c$ are the number of particles and the radius of the condensed state respectively. When close to gravitational collapse, $N_\rm{cond} \approx M_\rm{Cha}/m_\x$, where $M_\rm{Cha}$ is given in \eq{eq:M Cha}, and $r_c$ is given in \eq{eq:rcond}. 
It is then easy to verify the consistency of the description: Self-interactions render $r_c$ sufficiently big, such that the condition for diluteness becomes $\l_4 \ll 430$, which is of course also required for the perturbativity of the self-interaction, and is satisfied in all regions where bounds from neutron stars apply, as can be seen in Figs.~\ref{fig:NuChi}-\ref{fig:Coann} below.} 
Gravitational collapse is possible only after DM has condensed~\cite{Kouvaris:2012dz}, that is if
\beq
M_\rm{capt} > M_{_\rm{BEC}} = 3 \cdot 10^{42} \GeV \pare{\frac{m_\x}{\rm{GeV}}} \pare{\frac{T_\ns}{10^7 \: \rm{K} }}^3 \ .
\label{eq:M BEC}
\eeq

The size of the condensed state is determined in the early stages of condensation by the radius of the wavefunction of the ground state in the gravitational potential of the star~\cite{Kouvaris:2011fi} 
\beq
r_0 = \pare{\frac{8\p}{3} G \r_\ns m_\x^2}^{-1/4} \simeq 2 \cdot 10^{-6} \:\rm{m} \pare{\frac{\rm{GeV}}{m_\x}}^{1/2} .
\label{eq:r0}
\eeq
However, the BEC gravity can exceed the neutron star gravity if enough DM accumulates in the ground state, that is if $M_\rm{cond} = M_\rm{capt} - M_{_\rm{BEC}} > M_\rm{gr}$, where 
\beq
M_\rm{gr} 
\equiv \r_\ns \, \frac{4\p}{3} r_0^3 
\simeq 2 \cdot 10^{28} \GeV \pare{\frac{\rm{GeV}}{m_\x}}^{3/2}
\  .
\eeq
Then the size of the condensed state is determined by the BEC gravity and the DM self-interactions. The BEC gravity and self-interactions also determine the Chandrasekhar limit for gravitational collapse, and thus the fate of the condensate, as we shall now discuss.

\section{Gravitational collapse}
\label{sec:collapse}

A gas of bosons in its ground state can be withheld from gravitational collapse due to the uncertainty principle.\footnote{In order for a collection of particles to reach its ground state, some non-gravitational interaction -- either among the particles themselves, or between the particles and a surrounding heat bath (as is the case in the scenario under consideration) -- is necessary. In the absence of any non-gravitational interaction, a self-gravitating gas of particles will gravothermally disperse rather than condense and collapse~\cite{BinneyTremaine}.}
If the bosons have no self-interactions, equilibrium exists up to total mass~\cite{Ruffini:1969qy} 
\beq M_\rm{Cha} = 2 \mp^2/\p m_\x \ . \label{eq:MCha l=0}\eeq
Gravitational collapse then requires\footnote{Note that a gas of particles whose total mass exceeds the Chandrashekhar limit but which is not in its ground state, cannot collapse into a black hole. When not in the ground state, the particle momentum is not determined by the uncertainty principle for bosons, or the Pauli exclusion principle for fermions. The average particle momentum is larger, and provides additional pressure. The Chandrashekhar limit is then irrelevant. For a more detailed discussion, 
see Ref.~\cite{Kouvaris:2012dz}.}

\beq
M_\rm{cond} > 10^{38} \GeV \pare{\frac{\rm{GeV}}{m_\x}}  \  .
\label{eq:Mcapt>M Cha, l=0}
\eeq

In the presence of self-interactions, the equilibrium state is modified, and the new equilibrium conditions depend on the nature of the self-interactions.
For a repulsive contact-type self-interaction of a scalar field $\x$, described by the potential
$V_\rm{self} = \l_4 |\x|^4 /4$, the maximum mass for a stable configuration is
$M_\rm{Cha} = \frac{2 \mp^2}{\p m_\x} \pare{1 +\frac{\l_4}{32 \p} \frac{\mp^2}{m_\x^2}}^{1/2}$~\cite{Colpi:1986ye}.
This can be generalised for polynomial interactions of higher order~\cite{Ho:1999hs}
\bea
V_\rm{self} &=& \frac{\l_n}{n} |\x|^n \ ,
\label{eq:Vself}
\\
M_\rm{Cha} &\approx& \frac{2 \mp^2}{\p m_\x} \sqpare{1 +\frac{\l_n}{8 \p n} \frac{\mp^{n-2}}{m_\x^2}}^\frac{1}{n-2}  \  ,
\label{eq:M Cha}
\eea
for $n$ even, where $\l_n$ has mass dimension $4-n$. Equation~\eqref{eq:Vself} does not exhaust the range of possibilities for bosonic self-interactions -- importantly, it does not encompass the common case of self-interaction mediated by a heavy gauge boson. Nevertheless, we may regard \eq{eq:M Cha} as a reasonable approximation in the case of short-range repulsive interactions involving $n$ particles, provided that an appropriate correspondence between $\l_n$ and the self-coupling of interest is adopted, based on the amplitude of self-scattering.\footnote{In the case of long-range self-interaction, the functional dependence of $M_\rm{Cha}$ on the self-coupling is expected to be dramatically different (see e.g.~\cite{Jetzer:1989av}).}
The constraints on the DM self-interaction we derive in the following sections apply to the parameter
\beq
\l_{\x\x} \equiv 
32 \p \sqpare{\frac{\l_n m_\x^{n-4}}{8 \p n} }^\frac{2}{n-2}  
\label{eq:self coupling}
\eeq
(which coincides with $\l_4$ for $n=4$). When 
\beq 
\l_{\x\x} 
> 32 \p m_\x^2/\mp^2 
\simeq  6 \cdot 10^{-37} \pare{\frac{m_\x}{\rm{GeV}}}^2  \ , 
\label{eq:l domin} 
\eeq 
the self-interaction contribution dominates in \eq{eq:M Cha}, and the requirement for gravitational collapse becomes
\beq
M_\rm{cond} > 10^{56} \GeV  
\times
\l_{\x\x}^{1/2}
\pare{\frac{\rm{GeV}}{m_\x}}^2 \ . 
\label{eq:M Cha bound}
\eeq

\smallskip

The size of the condensed state when $M_\rm{cond}$ approaches $M_\rm{Cha}$ is~\cite{Colpi:1986ye,Ho:1999hs}
\beq
r_c \approx
\frac{1}{m_\x} \sqpare{1+ \frac{\l_n}{\p n} \, \frac{\mp^{n-2}}{m_\x^2}}^\frac{1}{n-2} 
\ .
\label{eq:rcond}
\eeq
Equation \eqref{eq:rcond} asserts that in the absence of self-interaction and when $M_\rm{cond} \approx M_\rm{Cha}$, the condensed particles are quasi-relativistic, $p_\x \sim 1/r_c \simeq m_\x$. This is in fact what gives rise to the Chandrashekhar limit: In the non-relativistic regime, the gravitational attraction, $V = -G M_\rm{cond} m_\x/r$, can always be balanced by the pressure due to the zero-point energy, $E_0 = p_\x^2/2m_\x \sim 1/2 m_\x r^2$, by $r$ becoming sufficiently small. However, when particles become relativistic, then $E_0 \simeq p_\x \sim 1/r$, and the kinetic pressure cannot balance the gravitational pressure if the total mass $M_\rm{cond}$ exceeds the Chandrachekhar limit of \eq{eq:MCha l=0}.
However, if the inequality \eqref{eq:l domin} holds, then \eq{eq:rcond} implies that the condensed particles are non-relativistic, $p_\x \sim 1/r_c < m_\x$. In this case, the gravitational pressure is balanced by the pressure due to the repulsive self-interactions, rather than the zero-point energy, up to the Chandrashekhar limit of \eq{eq:M Cha}. Note that $r_c$ of \eq{eq:rcond} is about equal to the Schwarchild radius of a black hole of mass $M_\rm{Cha}$, given by \eq{eq:M Cha}, $R_S = 2 M_\rm{Cha}/\mp^2$.

\medskip

If bosonic DM has attractive ($\l_{\x\x} <0$) rather than repulsive ($\l_{\x\x} >0$) self-interactions, the condensed DM particles inside the neutron star can coalesce and form solitonic bound states, known as $Q$-balls~\cite{Lee:1974ma,Coleman:1985ki}. This possibility arises particularly in SUSY theories~\cite{Kusenko:1997ad,*Kusenko:1997zq,Kusenko:1997si}, whose FDs of the scalar potential are often lifted to lowest order by (SUSY-breaking) attractive interactions.  Because of the formation of bound states, the dynamics of collapse change. We shall not consider the case of attractive self-interactions here.\footnote{Constraints on baryonic $Q$-ball DM from capture in neutron stars have been discussed in Ref.~\cite{Kusenko:2005du}. Those constraints, however, relate to $Q$-balls synthesised in the early universe rather than in the interior of a star.}

\section{Black hole accretion and Hawking evaporation}
\label{sec:BHaccretion}

If the DM captured in the neutron star condenses and exceeds the critical mass for gravitational collapse, it can form a black hole. The black hole accretes mass from the star and can potentially destroy it. However, the growth of the black hole can be stalled by Hawking evaporation. The accretion rate onto the black hole is proportional to $M_\rm{BH}^2$, while the evaporation rate is proportional to $M_\rm{BH}^{-2}$, where $M_\rm{BH}$ is the black hole mass. Whether the black hole will grow and destroy the star, or evaporate with no observable consequences, thus depends on the balance between accretion and evaporation at the time of the black hole formation.
The growth of the black hole is governed by the equation
\beq
\frac{dM_\rm{BH}}{dt} = 
 \left.\frac{dM_\rm{BH}}{dt}\right|_\rm{NS}
+\left.\frac{dM_\rm{BH}}{dt}\right|_\rm{DM}
+\left.\frac{dM_\rm{BH}}{dt}\right|_\rm{H} 
\ .
\label{eq:BH growth}
\eeq
We now discuss the various contributions.

\setcounter{paragraph}{0}
\paragraph{Accretion of neutron star matter.} It can be estimated in the hydrodynamic spherical approximation (Bondi regime)
\beq
\left.\frac{dM_\rm{BH}}{dt}\right|_\rm{NS} = \frac{4 \p \l \, \r_\ns \, G^2 M_\rm{BH}^2}{c_s^3} \ ,
\label{eq:accretion NS}
\eeq
where $\l = 0.25$ and $c_s \simeq 0.17$ is the speed of sound inside the star~\cite{ShapiroTeukolsky}.

\paragraph{Accretion of dark matter.} It includes two contributions: the accretion from the condensed DM component, and the accretion from the thermalised DM in the excited states. (The accretion of DM from the galactic halo is negligible.)

The accretion of (non-condensed) thermal DM particles can be estimated also in the hydronamic spherical approximation. Even in the limit of zero self-interactions, DM particles are not collisionless; their collisions with neutrons keep them in thermal equilibrium within the star. The collisions are possible, even if the occupation number of the thermal levels is saturated, since the thermal DM particles can lose energy by moving to the ground state. In the hydrodynamic approximation, the thermal DM accretion rate is given by \eq{eq:accretion NS}, with $\r_\ns$ replaced by the density of the thermal DM particles, $\r_{\x, \rm{thermal}}$.

If the accretion rate of thermal DM particles onto the black hole,
$4 \p \l \, \r_{\x, \rm{thermal}} \, G^2 M_\rm{BH}^2  / c_s^3$, is smaller than the DM capture rate in the star, then the thermal states remain always filled and $\r_{\x, \rm{thermal}} = m_\x n_{_\rm{BEC}}$. 
In this case, the newly captured DM particles are added to the ground state and contribute to $\r_\rm{cond}$. However, if the initial mass of the black hole is large enough, then the accretion rate of thermal DM particles to the black hole may exceed the DM capture rate in the star. In this case, $\r_{\x, \rm{thermal}} < m_\x n_{_\rm{BEC}}$, and no particles reside in the condensed state, $\r_\rm{cond} = 0$. This possibility arises in the presence of DM self-interactions, which increase the Chandrashekhar limit and thus the initial mass of the black hole.

The extent of the wavefunction of the newly condensed DM particles is determined by the gravitational field of the black hole, which dominates over that of the neutron star in the relevant region. For the $\propto 1/r$ potential of the black hole, this is $a_0 \sim \mp^2/(M_\rm{BH} m_\x^2)$, which implies that $a_0 < R_S = 2 M_\rm{BH}/\mp^2$ (since $M_\rm{BH} \geqslant \mp^2/m_\x$, due to the Chandrashekhar limit of \eq{eq:MCha l=0}). Thus, the rate at which DM particles are added to the ground state is also their rate of accretion to the black hole.

\paragraph{Hawking evaporation.} The rate is given by
\beq
\left.\frac{dM_\rm{BH}}{dt}\right|_\rm{H} = -\frac{1}{15360 \p G^2 M_\rm{BH}^2} \  .
\label{eq:Hawking}
\eeq
As the black hole evaporates, it will produce light SM particles, as well as light dark-sector particles. The latter are expected to exist in the ADM scenario because of the necessity to efficiently annihilate away the symmetric component of DM in the early universe. The radiation emitted is not sufficient to significantly heat the nucleons in the neutron star, even if it is emitted entirely in the form of SM particles~\cite{McDermott:2011jp}. Thus the accretion of neutron star matter onto the black hole remains as described above. However, if some of the Hawking radiation is emitted in the form of dark-sector particles which interact strongly with DM, it is possible that the DM residing in the neutron star is heated significantly. This could greatly reduce, if not eliminate, the DM accretion onto the black hole~\cite{McDermott:2011jp}.  Because the accretion of DM onto the black hole depends on assumptions about the dark sector, we will ignore this contribution. This leads to conservative bounds.

Equations \eqref{eq:BH growth}, \eqref{eq:accretion NS} and \eqref{eq:Hawking} imply that accretion onto the black hole overpowers Hawking evaporation, and the black hole grows, if the black hole is sufficiently massive at its birth, $M_{_\rm{BH}} > M_\rm{evap}$, where
\beq
M_\rm{evap} 
\equiv  \pare{\frac{c_s^3 \, \mp^8}{61440 \p^2 \l \r_\ns}}^{1/4} 
\simeq 5 \cdot 10^{36} \GeV \  .
\label{eq:Mevap}
\eeq
%

\section{Bounds}
\label{sec:limits}

The above considerations imply that observations of neutron stars can exclude the regions of parameter space of ADM models which satisfy~\cite{Kouvaris:2011fi}
\bea
M_\rm{capt}	-	M_{_\rm{BEC}} \geqslant	M_\rm{Cha} \geqslant	M_\rm{evap}  \ ,
\label{eq:excl reg}
\eea
and as long as the inequalities \eqref{eq:therm condition} and \eqref{eq:thermal evap} hold. Here $M_\rm{capt}$, $M_{_\rm{BEC}}$, $M_\rm{Cha}$ and $M_\rm{evap}$ are given by \eqs{eq:Mcapt num}, \eqref{eq:M BEC}, \eqref{eq:M Cha} and \eqref{eq:Mevap} respectively. 
The most constraining bounds arise from old neutron stars with low core temperatures, located in DM-dense regions.

The core temperature $T_\ns$ of any neutron star (in the region $r \lesssim 0.9 R_\ns$) cannot be measured directly; it must be related to the surface temperature
$T_{_\text{NS, surf}}$ via a thermal conduction model. Adopting the standard, two-zone, heat-blanket model, one arrives at the empirical relation~\cite{Gudmundsson:1982}
\beq
 T_\ns  = 2.0 \times 10^6 \pare{ \frac{T_{_\text{NS, surf}} }{ 10^5 \, \rm{K}} }^{1.8}~.
\eeq
Unfortunately, thermal emission from old, cold neutron stars is rarely observed. 
At the time of writing, there exists a single reliable measurement of $T_{_\text{NS, surf}}$ for a recycled object older than $\sim 10^9 \yr$, namely PSR J0437$-$4715, whose ultraviolet thermal flux 
has been fitted by various spectral models to give $T_{_\text{NS, surf}} = 10^{5.1\pm0.1} \, \rm{K}$~\cite{Kargaltsev:2003eb}. Robust upper limits on the thermal flux have also been obtained for the millisecond pulsars PSR J0030$+$0451 and PSR J2124$-$3358, implying $T_{_\text{NS, surf}} \leq 10^{6.0}\,{\rm K}$ and $10^{5.7}\,{\rm K}$ respectively~\cite{Koptsevich:2002dg,*Mignani:2003nw}.
Younger objects with ages $\sim 10^7 \yr$ exhibit comparable upper limits, suggesting that reheating occurs as the star ages, e.g. via the rotochemical mechanism, to keep $T_{_\text{NS, surf}}$ above $\sim 10^5\,\rm{K}$~\cite{Kargaltsev:2003eb,Reisenegger:1994be}. For example, PSR B1929$+$10 and PSR B0950$+$08 have $T_{_\text{NS, surf}} \leq 10^{6.0}\,{\rm K}$ and $10^{5.7}\,{\rm K}$ respectively~\cite{Mignani:2002er,*Pavlov:1996gw}. In all these cases, $T_{_\text{NS, surf}}$ and hence $T_\ns$ are subject to a range of uncertainties surrounding the many possible choices of spectral model (e.g. absorbed blackbody, absorbed power law), geometrical factors (e.g. polar cap emission), and compositional effects~\cite{Zavlin:2001ga}.

\begin{figure*}[ht]
\includegraphics[width=0.9\columnwidth]{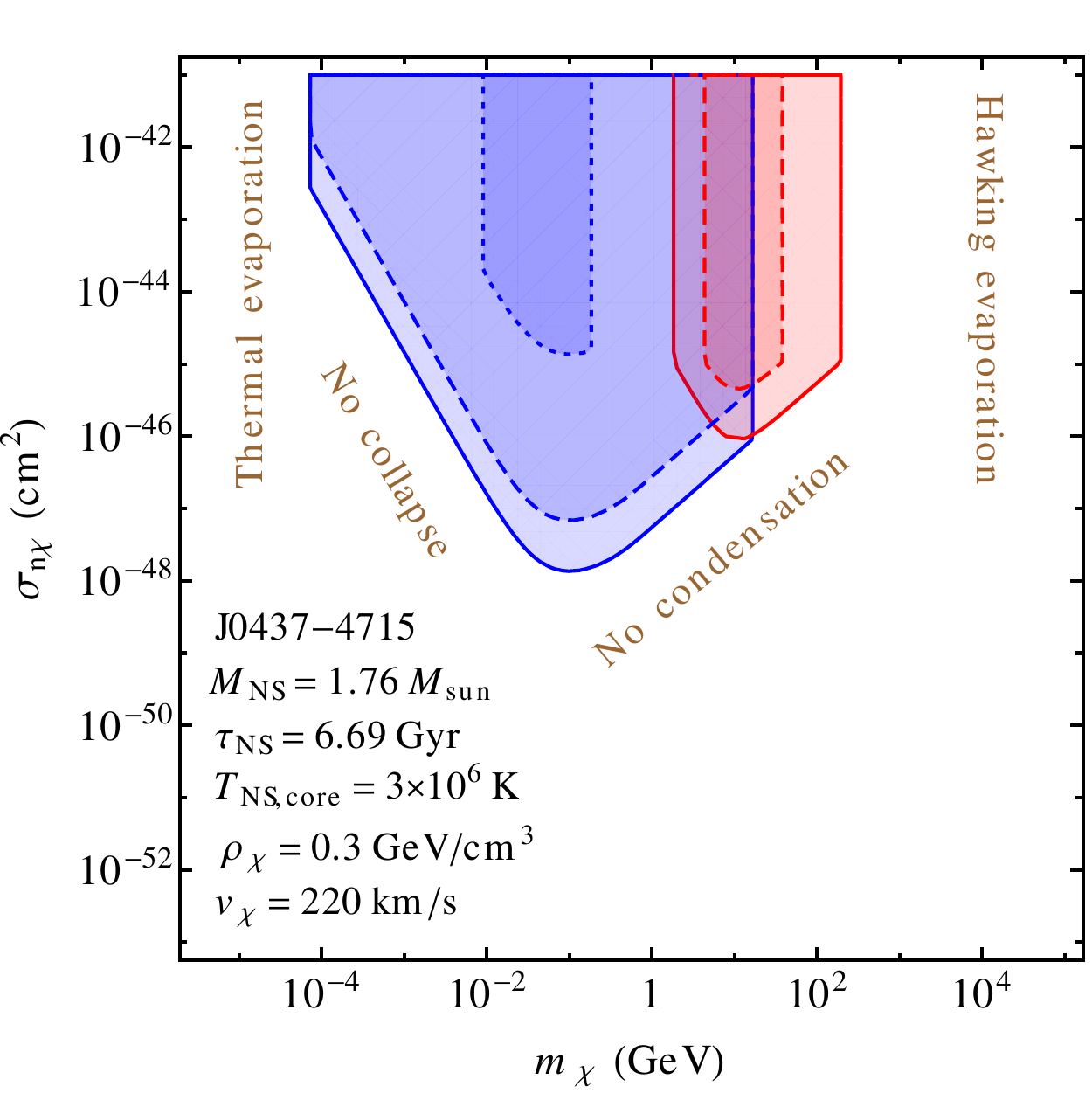}
\hspace{0.1\columnwidth}
\includegraphics[width=0.9\columnwidth]{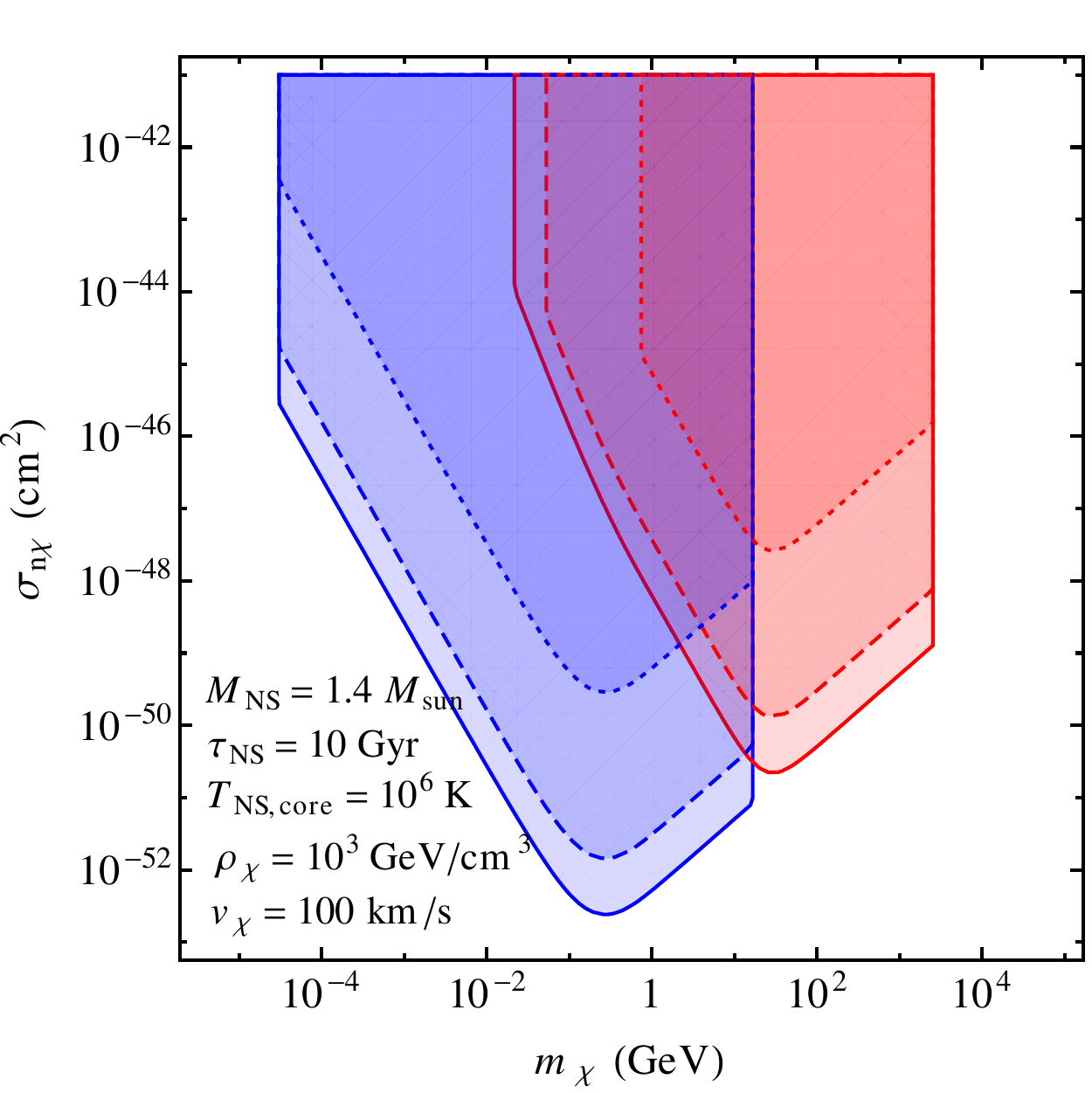}
\caption{
Bounds on the DM-nucleon cross-section $\s_{n\x}$, for DM self-interaction strengths $\l_{\x\x} =  0$ (blue-shaded regions to the left of each plot) and $\l_{\x\x} = 10^{-25}$ (red-shaded regions to the right of each plot), and for DM-nucleon co-annihilation cross-sections $\sva = 0$ (solid lines), $\sva = 5 \cdot 10^{-56} \cm^3/\snd$ (dashed lines) and $\sva=10^{-53} \cm^3/\snd$ (dotted lines). Shaded areas are disfavoured for the astrophysical parameters shown on the plots. 
}
\label{fig:NuChi}
\end{figure*}

\begin{figure*}[ht]
\includegraphics[width=0.9\columnwidth]{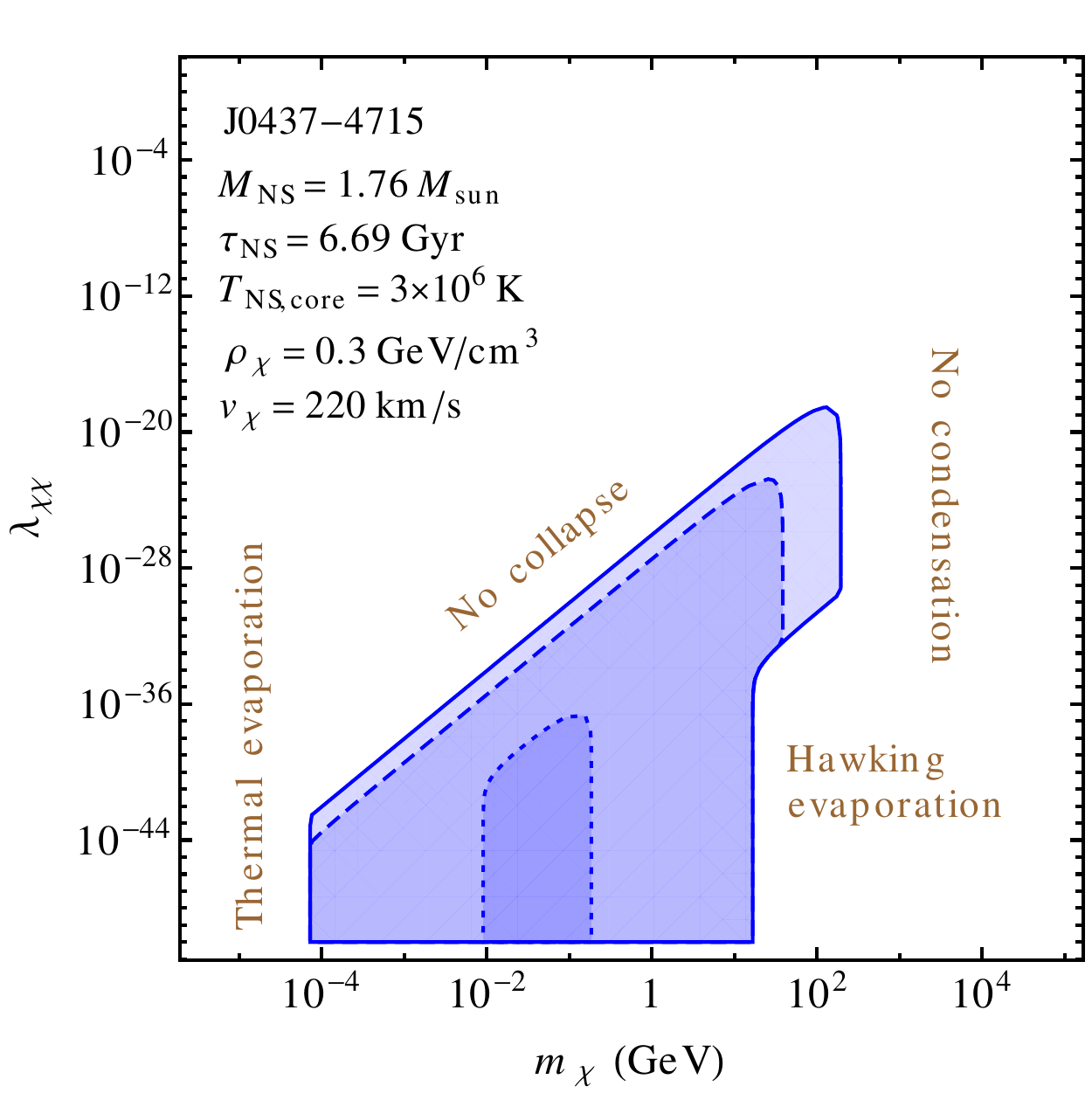}
\hspace{0.1\columnwidth}
\includegraphics[width=0.9\columnwidth]{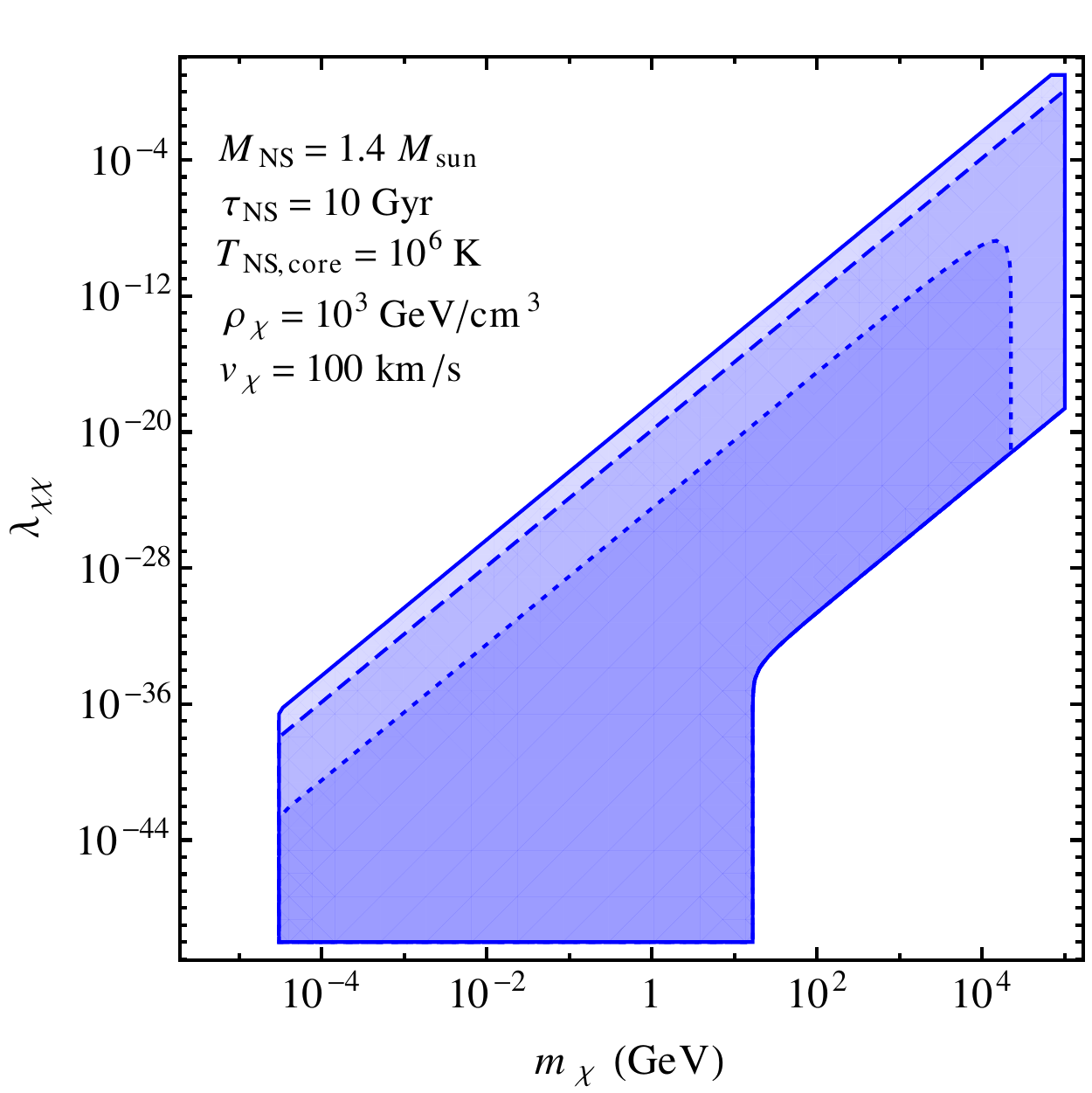}
\caption{ 
Bounds on the DM self-coupling $\l_{\x\x}$, defined in \eq{eq:self coupling}, for $\s_{n\x} \geq \s_\rm{sat}$ (i.e. maximum DM capture in the neutron star, for the entire DM mass range considered), and for DM-nucleon co-annihilation cross-sections $\sva = 0$ (bounded by the solid lines), $\sva = 5 \cdot 10^{-56} \cm^3/\snd$ (bounded by the dashed lines) and $\sva=10^{-53} \cm^3/\snd$ (bounded by the dotted lines). 
Shaded areas are disfavoured for the astrophysical parameters shown on the plots.}
\label{fig:Self}
\end{figure*}

\begin{figure*}[ht]
\includegraphics[width=0.9\columnwidth]{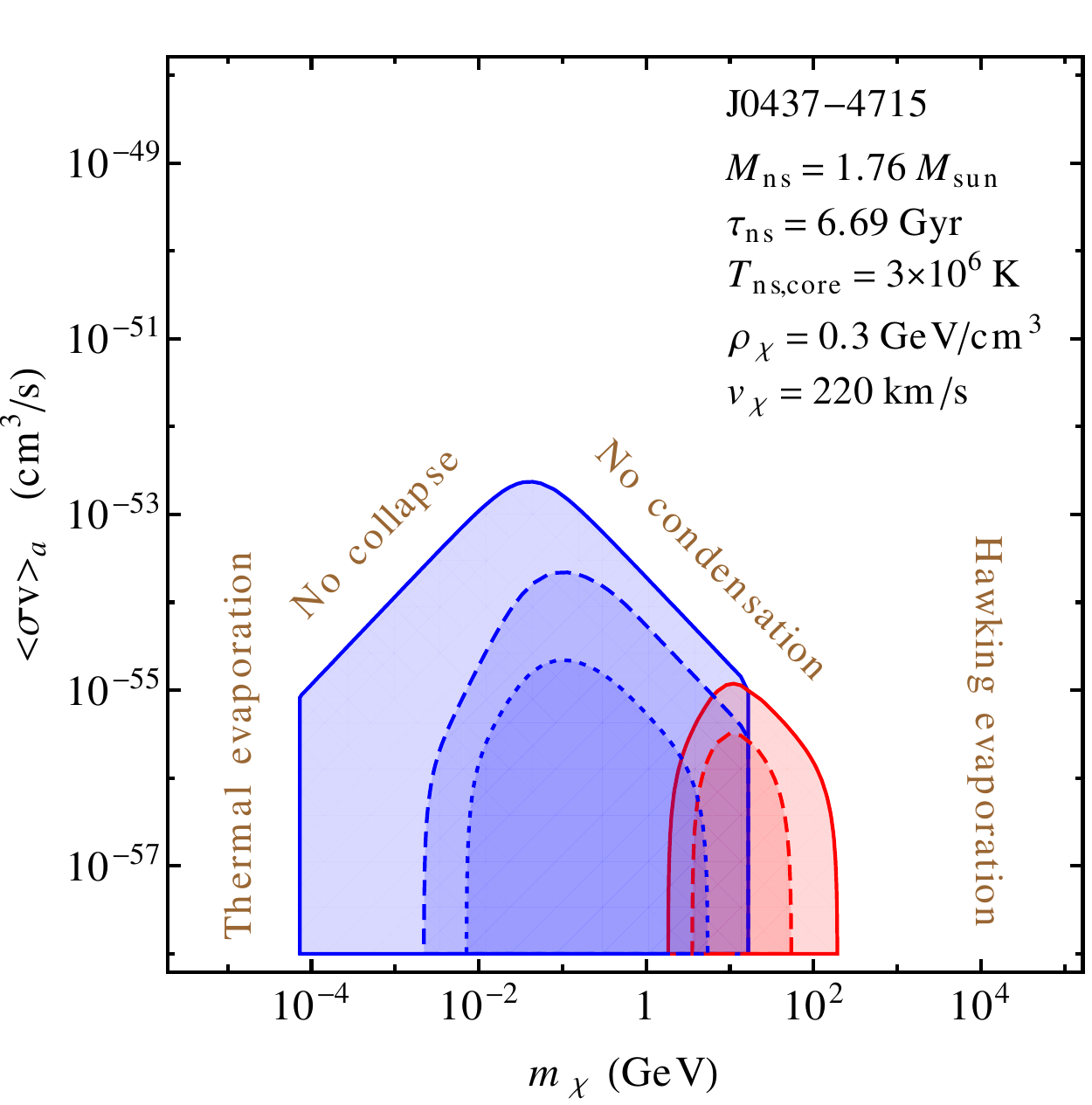}
\hspace{0.1\columnwidth}
\includegraphics[width=0.9\columnwidth]{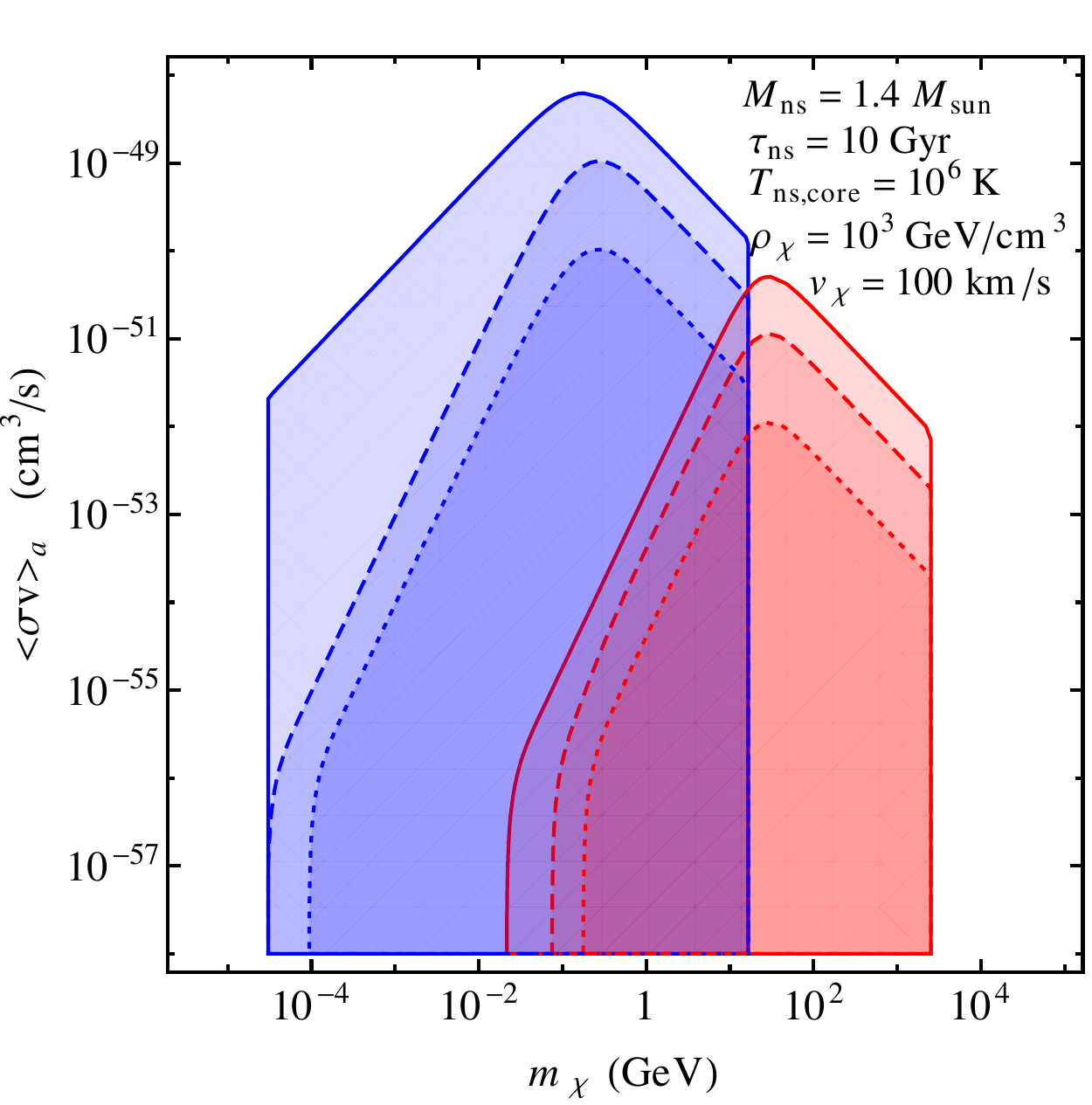}
\caption{
Bounds on the DM-nucleon coannihilation cross-section $\ang{\s v}_a$, for DM-nucleon scattering cross-section $\s_{n\x} \gtrsim \s_\rm{sat}$ (solid lines), $\s_{n\x} = 3 \cdot 10^{-46} \cm^2$ (dashed lines) and $\s_{n\x} = 3 \cdot 10^{-47} \cm^2$ (dotted lines), and for DM self-coupling $\l_{\x\x} = 0$ (blue-shaded regions to the left of each plot) and $\l_{\x\x} = 10^{-25}$ (red-shaded regions to the right of each plot). Shaded areas are disfavoured for the astrophysical parameters shown on the plots. 
}
\label{fig:Coann}
\end{figure*}

Let us begin by examining PSR J0437$-$4715, which has estimated age $\t_\ns  \approx 6.69 \Gyr$~\cite{Manchester:2004bp}, and the lowest $T_\ns$ known. It is a high-mass pulsar, with $M_\ns \approx 1.76 M_\odot$~\citep{Verbiest:2008gy}. Its measured thermal UV emission implies surface temperature $T_{_\text{NS, surf}} \approx 10^{5.1} \: \rm{K}$~\cite{Kargaltsev:2003eb}, which corresponds to core temperature $T_\ns \approx 3 \cdot 10^6 \: \rm{K}$~\cite{Gudmundsson:1982}. 
It is located about $160\kpc$ from our solar system~\cite{Manchester:2004bp}, and thus we take $\r_\x   = 0.3 \GeV/\cm^3, \: \bar{v}_\x  = 220 \km / \snd$. We present the constraints from PSR J0437-4715 on the DM-nucleon scattering, DM self-interaction and DM-nucleon co-annihilation, in the left plots of Figs.~\ref{fig:NuChi}-\ref{fig:Coann}.

In the right plots of Figs.~\ref{fig:NuChi}-\ref{fig:Coann} we consider the possibility of an old pulsar with low core temperature, located closer to the center of the galaxy, where the DM density is expected to be larger.
We adopt the parameters $ M_\ns = 1.4 M_\odot, 
\: R_\ns = 10 \km, 
\: T_\ns = 10^6 \: \rm{K} , 
\: \t_\ns  = 10 \Gyr, \: \r_\x   = 10^3 \GeV/\cm^3, \: \bar{v}_\x  = 100 \km / \snd$. These parameters lead to more stringent constraints than those derived for J0437-4715, albeit they do not correspond to an observed pulsar.

\section{Discussion and conclusions}
\label{sec:conc}

{\it Non-annihilating} dark matter in the form of fundamental bosonic particles having weak-scale interactions with ordinary matter remains a viable possibility. Although the existence of old neutron stars can set interesting limits on this scenario, these limits are highly sensitive to DM {\it self-interactions}.  We have discussed why self-interactions of sufficient strength to alter the neutron star limits are almost inevitable, and estimated their size in some representative models.  Furthermore, even if DM is strictly non-annihilating, {\it co-annihilations} with nucleons are possible and arise naturally in many asymmetric DM models.  Such co-annihilations cap the density of DM accumulated in neutron stars, again acting to prevent gravitational collapse.  The effect of both self-interactions and co-annihilations is quantified in Figs.~\ref{fig:NuChi}, \ref{fig:Self}, and \ref{fig:Coann}. 

The bounds we have derived by considering the pulsar J0437-4715 imply that:
\benu[(i)]
\item
Repulsive self-interactions shift the excluded mass interval towards higher values. This is because self-interactions increase the total mass required for
gravitational collapse, which can only be fulfilled by larger DM masses. Since the resulting black holes are more massive at formation, accretion becomes more effective and Hawking evaporation less effective.  This moves the upper limit on the excluded $m_\chi $ range to higher values, up to an ultimate value of $\sim 200 \GeV$. For larger masses, the number density of DM in the neutron star is not sufficient for condensation, and no limits apply. 

Thus, the entire DM mass range is allowed, even for saturated capture and no co-annihilations, if
\beq
\l_{\x\x} \ \gtrsim \ \left.\frac{(2\p)^3 M_\rm{capt}^2 m_\x^4}{ \mp^6}\right|_{m_\x \sim 200 \GeV} \sim  10^{-18} \: . 
\eeq
For a scalar quartic interaction, this corresponds to self-scattering cross-sections $\s_{\x\x} = \l_{\x\x}^2/64 \p m_\x^2$ greater than  
\beq \s_{\x\x} \gtrsim 10^{-70} \cm^2 \: . \eeq

The DM mass range below $\sim 10 \GeV$ -- preferred in the ADM scenarios and favoured by DM direct-detection experiments -- is allowed for even smaller self-couplings, $\l_{\x\x} \gtrsim 10^{-22}$ and   
$\s_{\x\x} \gtrsim 10^{-76} \cm^2$.

\smallskip

Evidently, the magnitude of $\s_{\x\x}$ necessary to evade the bounds is several decades of orders of magnitude smaller than the minimum $\s_{n\x}$ for which bounds apply, $\s_{n\x} \gtrsim 10^{-48} \cm^2$.
In view of the arguments of section~\ref{sec:self}, it is expected that the DM-nucleon coupling responsible for the DM capture in the star, will induce a much stronger DM self-interaction than what is necessary to evade the limits. The only possible exception to this is supersymmetric models with high scale of SUSY-breaking mediation.

\item  
If DM corresponds to a FD field in a SUSY theory, then the self-interaction induced by SUSY breaking (see \eq{eq:lambda susy}), if repulsive, un-excludes masses  $m_\x \lesssim \sqpare{\l_{\x\x} \mp^6 / (2\p)^3 M_\rm{capt}^2 }^{1/4}$, or
\begin{align}
&m_\x \lesssim&  \!\!\! & 40 \MeV \pare{\frac{m_s}{\TeV}}^{1/2},&  &\rm{PMSB}
\\
&m_\x \lesssim&  \!\!\! & 10 \GeV \sqpare{ \frac{m_s / \TeV}{M_m / (10^{14}\GeV) } }^{1/2},&  &\rm{GMSB} \ .
\end{align}
Here we adopted the current rough experimental lower limit $m_s \sim \rm{TeV}$ as an indicative value for the soft scale. Note however that in the case of GMSB, the soft scale in the dark sector may be higher or lower than in the ordinary sector, if SUSY breaking is mediated to the dark sector more directly or less directly respectively than it is mediated to the ordinary sector. 

The case of PMSB admits the most stringent bounds on bosonic ADM, though a significant mass range is still viable. In models with low-energy SUSY-breaking mediation, the range $m_\x \lesssim 10 \GeV$ is viable even for messenger scale as high as $10^{14} \GeV$.
\item
Co-annihilations shrink the excluded parameter space, and eliminate all constraints if the  DM-baryon co-annihilation is stronger than
\beq
\sva \gtrsim  10^{-52} \cm^3 / \snd \  .
\label{eq:sigmav a no bounds}
\eeq 
This is well within the range of values that appears in models in the recent literature~(see e.g. \cite{Davoudiasl:2011fj}).
\item
There is no exclusion for DM-nucleon scattering cross-sections lower than 
\beq 
\s_{n\x} \lesssim 10^{-48} \cm^2 
\  ,
\label{eq:sigma upper ultimate}
\eeq
even for vanishing self-interactions and no co-annihilations.
Note that the success of the ADM scenario does not rely on weak-scale interactions of DM with ordinary matter, as in the case in the thermal relic DM scenario (also known as the WIMP miracle). Thus Eq.~\eqref{eq:sigma upper ultimate} is possible (though, unfortunately, very small $\s_{n\x}$ would preclude the possibility of DM direct detection).

\eenu

The above numerical values refer specifically to bounds from J0437-4715, which are likely the most stringent bounds that can be derived based on observation. However, the conclusions remain qualitatively the same for other plausible sets of astrophysical parameters, such as the one used in the right-side plots of Figs.~\ref{fig:NuChi}--\ref{fig:Coann}. The bounds rescale according to the equations presented in the paper. Note that the bounds depend sensitively on $T_\ns$, and for large DM masses, $m_\x \gtrsim 100 \GeV$, most neutron stars' core is too hot to provide useful constraints.

Lastly, we note that the constraints for vanishing DM self-interaction do not apply to the case of attractive self-interactions. Constraining the latter possibility requires taking into account the dynamics of bound-state formation.

\subsection*{Uncertainties in the bounds of Figs.~\ref{fig:NuChi}-\ref{fig:Coann}.}

\setcounter{paragraph}{0}
\paragraph{Dark-matter capture.}

The amount of DM captured in the neutron star depends on the DM density and average velocity in the vicinity of the star (see \eq{eq:Mcapt num}), whose precise values are of course unknown. 
Because the dependence on the local DM density is linear, the various fluctuations due to non-spherical DM density distribution in the halo and possible local overdensities and underdensities, are expected to average out along the trajectory of the star in the galaxy. The uncertainty of the average DM density in the solar region, where PSR J0437$-$4715 is located, is thought to be around 30\%~\cite{Salucci:2010qr,*Bovy:2012tw}.
The DM velocity dispersion is approximated by the rotational velocity of a spherical halo at the relevant distance. The observed triaxiality of haloes introduces a variation of the order of 50\% in the velocity dispersion~\cite{Oguri:2010ik}. 
Thus, the uncertainty in the determination of $M_\rm{capt}$ due to imprecise knowledge of the local DM density and average velocity is estimated to be less than one order of magnitude.
The lines marked as ``no collapse" and ``no condensation" in Figs.~\ref{fig:NuChi}, \ref{fig:Coann} and in Fig.~\ref{fig:Self} depend linearly and quadratically on this quantity, respectively. They are thus expected to be good estimates within one and two orders of magnitude respectively.

Note that the capture radius, $GM_\ns/v^2 \sim  10^{-7} \pc$, is small compared to the length scale of the halo, so the background velocity distribution function is isotropic to a good approximation in the vicinity of the compact object. Also, the capture rate estimation of \eq{eq:C nx} includes the gravitational and loss cone modifications (the latter producing an anisotropic velocity distribution function).

\paragraph{Accretion on the black hole.}

In Sec.~\ref{sec:BHaccretion}, the accretion of neutron-star matter onto the black hole was estimated in the hydrodynamic spherical approximation, ignoring the rotation of the star. However, infalling matter has to dissipate its angular momentum before it can be accreted onto the black hole. This can occur via formation of an accretion disk around the black hole, and transport of the angular momentum to its outer regions via viscous processes. The accretion rate can then be estimated in the Shakura-Sunyaev disk model~\cite{Shakura:1972te}, with the viscosity of the infalling matter estimated by the standard alpha prescription (proportional to the gas and/or magnetic pressure) or from Landau-damped electron and neutron scattering~\cite{Cutler:1987,*Shternin:2008es}. The effect will be to reduce the accretion rate of neutron matter onto the black hole with respect to the estimate of Sec.~\ref{sec:BHaccretion}. This will shift the point of balance between accretion and Hawking evaporation, moving
the rightmost vertical boundaries in Figs.~\ref{fig:NuChi}-\ref{fig:Coann}, to lower excluded mass and shrinking the excluded region. In that sense, the omission of disk accretion is conservative. Detailed estimation of the effect of rotation on the fate of the black hole is beyond the scope of this work. 

We note however that, high angular momentum accretion via a viscous accretion disk can be accelerated substantially in the vicinity of a compact object,
if the disk is warped by general relativistic or radiative forces, both of which are likely to affect the system under consideration; simulations suggest that the accretion rate can increase $10^2$-fold or more~\cite{Nixon:2012xx,*Nixon:2012nx}.

Accretion will also be modified by magnetic stresses arising from the neutron star's internal magnetization, both the fossilized component and that wound up by the rotating, infalling matter. Magnetic stresses can exceed viscous stresses and introduce complicated topological issues, which are potentially significant but lie well outside the scope of this paper.

\paragraph{Hawking evaporation.}
The black hole itself inherits the angular momentum of the collapsed matter. The spinning of the black hole can affect the rate of its evaporation, which in \eq{eq:Hawking} was estimated for  the non-rotating case. For a rotating black hole, the rate of evaporation is suppressed with respect to \eq{eq:Hawking} by the factor~\cite{Murata:2006pt}
\beq  2  \pare{1+\frac{1}{\sqrt{1 - J^2 \mp^4/M_\rm{BH}^4 }}}^{-1} \ , 
\label{eq:Hawking suppresion}
\eeq
where $J$ is the angular momentum of the black hole. We may estimate the angular momentum of the newly formed black hole inside the neutron star as 
\beq  J \sim \pare{\frac{2}{5}M_\rm{Cha} r_c^2} 2\p f_\ns \label{eq:J} \eeq
where $f_\ns$ is the frequency of rotation of the neutron star, and for J0437$-$4715, $f_\ns^{-1} \simeq 5.8 \, \rm{ms}$~\cite{Manchester:2004bp}. Using \eqs{eq:M Cha}, \eqref{eq:rcond}, we find that $J \mp^2/M_\rm{Cha}^2 \ll 1$ in all of the excluded parameter space, thus the effect of rotation on the evaporation of the black hole can be safely neglected.

\medskip
\noindent
\textbf{Note added:} As we were finalising this manuscript, Ref.~\cite{Bramante:2013hn} appeared on the arXiv. In that paper, the authors derive constraints for the case of self-interacting bosonic ADM, and they also consider the effect of DM self-annihilations. Here we have discussed in detail when and why DM self-interactions are inevitable, and we have considered the effect of DM-nucleon co-annihilations, motivated by the generic symmetry structure of ADM models.

\section*{Acknowledgements}

We thank Ray Volkas, Robert Foot, Andy Martin, Michael Schmidt and Marieke Postma for useful discussions. NFB, AM and KP were supported by the Australian Research Council. KP was also supported by the Netherlands Foundation for Fundamental Research of Matter (FOM) and the Netherlands Organisation for Scientific Research (NWO).


\bibliography{BH_arXiv_v3.bbl}

\end{document}